\documentclass[twocolumn,twocolappendix]{aastex631}

\usepackage{amsmath}
\usepackage{empheq}
\usepackage{mathrsfs}
\usepackage{textcomp}
\usepackage{enumitem}   
\usepackage{gensymb}
\usepackage{hyperref}
\usepackage{graphicx}
\usepackage[caption=false]{subfig}
\usepackage{multirow}
\usepackage{longtable}
\usepackage[flushleft]{threeparttable}
\usepackage{booktabs} 
\usepackage{CJK}
\hypersetup{colorlinks, linkcolor={blue}, citecolor={blue}, urlcolor={blue}} 

\definecolor{DarkOrange}{RGB}{204, 85, 0}
\definecolor{LincolnGreen}{RGB}{17, 102, 0}
\def\ion#1#2{#1$\;${\footnotesize\rm{#2}}\relax}

\usepackage{xspace}

\newcommand\nicer{\textit{NICER}\xspace}

\newcommand\swift{\textit{Swift}\xspace}
\newcommand\maxi{\textit{MAXI}\xspace}
\newcommand\srg{\textit{SRG}\xspace}

\newcommand\rosat{\textit{ROSAT}\xspace}
\newcommand\gaia{\textit{Gaia}\xspace}
\newcommand\target{AT2019wey\xspace}

\def \caltech {{Division of Physics, Mathematics and Astronomy, 
California Institute of Technology, Pasadena, CA 91125, USA}}

\def \coo {{Caltech Optical Observatories, California Institute of Technology, Pasadena, CA 91125, USA}}

\def \gsfc {{Astrophysics Science Division, NASA Goddard Space Flight Center, Greenbelt, MD 20771, USA}}

\shorttitle{AT2019wey: a candidate BH LMXB}
\shortauthors{Yao et al.}

\begin{document}

\title{Multi-wavelength Observations of AT2019wey: \\
a New Candidate Black Hole Low-mass X-ray Binary}

\correspondingauthor{Yuhan Yao}
\email{yyao@astro.caltech.edu}

\author[0000-0001-6747-8509]{Yuhan Yao}
\affiliation{\caltech}

\author[0000-0001-5390-8563]{S. R. Kulkarni}
\affiliation{\caltech}


\author[0000-0002-7226-836X]{Kevin B. Burdge}
\affiliation{\caltech}

\author[0000-0002-4770-5388]{Ilaria Caiazzo}
\affiliation{\caltech}

\author[0000-0002-8989-0542]{Kishalay De}
\affiliation{\caltech}

\author[0000-0001-9584-2531]{Dillon Dong}
\affiliation{\caltech}

\author[0000-0002-4223-103X]{C. Fremling}
\affiliation{\caltech}

\author[0000-0002-5619-4938]{Mansi M. Kasliwal}
\affil{\caltech}

\author[0000-0002-6540-1484]{Thomas Kupfer}
\affiliation{Texas Tech University, 
    Department of Physics \& Astronomy,
    Box 41051, 79409, Lubbock, TX, USA}

\author[0000-0002-2626-2872]{Jan van~Roestel}
\affiliation{\caltech}

\author[0000-0003-1546-6615]{Jesper Sollerman}
\affiliation{The Oskar Klein Centre, Department of Astronomy, 
	Stockholm University, AlbaNova, SE-10691 Stockholm, Sweden}


\author{Ashot Bagdasaryan}
\affiliation{\caltech}

\author[0000-0001-8018-5348]{Eric C. Bellm}
\affiliation{DIRAC Institute, Department of Astronomy, 
        University of Washington, 
        3910 15th Avenue NE, Seattle, WA 98195, USA}
        
\author[0000-0003-1673-970X]{S. Bradley Cenko}
\affiliation{\gsfc}
	
\author{Andrew J. Drake}
\affiliation{\caltech}
	
\author[0000-0001-5060-8733]{Dmitry A. Duev}
\affiliation{\caltech}
	
\author[0000-0002-3168-0139]{Matthew J. Graham}
\affiliation{\caltech}

\author{Stephen Kaye}
\affiliation{\coo}

\author[0000-0002-8532-9395]{Frank J. Masci}
\affiliation{IPAC, California Institute of Technology, 1200 E. California
             Blvd, Pasadena, CA 91125, USA}
             
\author[0000-0001-6070-7540]{Nicolas Miranda}
\affiliation{Institut f\"{u}r Informatik, Humboldt-Universit\"{a}t zu Berlin, Rudower Chaussee 25, 12489 Berlin, Germany}
             
\author[0000-0002-8850-3627]{Thomas A. Prince}
\affiliation{\caltech}

\author[0000-0002-0387-370X]{Reed Riddle}
\affiliation{\coo}

\author[0000-0001-7648-4142]{Ben Rusholme}
\affiliation{IPAC, California Institute of Technology, 1200 E. California
             Blvd, Pasadena, CA 91125, USA}
             
\author[0000-0001-6753-1488]{Maayane T. Soumagnac}
\affiliation{Lawrence Berkeley National Laboratory, 
    1 Cyclotron Road, Berkeley, CA 94720, USA}
\affiliation{Department of Particle Physics and Astrophysics, 
    Weizmann Institute of Science, Rehovot 76100, Israel}

\begin{abstract}
AT2019wey (SRGA\,J043520.9$+$552226, SRGE\,J043523.3$+$552234) is a transient first reported by the ATLAS optical survey in 2019 December. It rose to prominence upon detection, three months later, by the \textit{Spektrum-Roentgen-Gamma} (\srg) mission in its first all-sky survey. X-ray observations reported in \citet{Yao2020} suggest that AT2019wey is a Galactic low-mass X-ray binary (LMXB) with a black hole (BH) or neutron star (NS) accretor. Here we present ultraviolet, optical, near-infrared, and radio observations of this object. 
We show that the companion is a short-period ($P\lesssim16$\,hr) low-mass ($<1\,M_\odot$) star. 
We consider AT2019wey to be a candidate BH system since its locations on the $L_{\rm radio}$--$L_{\rm X}$ and $L_{\rm opt}$--$L_{\rm X}$ diagrams are closer to BH binaries than NS binaries. 
We demonstrate that from 2020 June to August, despite the more than 10 times brightening at radio and X-ray wavelengths, the optical luminosity of \target only increased by 1.3--1.4 times. 
We interpret the UV/optical emission before the brightening as thermal emission from a truncated disk in a hot accretion flow and the UV/optical emission after the brightening as reprocessing of the X-ray emission in the outer accretion disk. AT2019wey demonstrates that combining current wide-field optical surveys and \srg provides a way to discover the emerging population of short-period BH LMXB systems with faint X-ray outbursts. 
\end{abstract}

\section{Introduction}
\subsection{Low-mass X-ray Binaries and \srg}
Low-mass X-ray binaries (LMXBs) contain an accreting neutron star (NS) or black hole (BH) in orbit with a low-mass ($\lesssim 2\,M_\odot$) companion star. Most of the known BH LMXBs were discovered by X-ray all-sky monitors (ASMs) during X-ray outbursts induced by instabilities in the accretion processes. The most sensitive X-ray ASM to date, the Monitor of All-sky X-ray Image (\maxi; \citealt{Matsuoka2009}), has a transient triggering threshold of $8$\,mCrab ($\rm 1\,mCrab = 2.4\times 10^{-11}\,{\rm erg\,s^{-1}\, cm^{-2}}$ over 2--10\,keV) sustained for 4\,days \citep{Negoro2016}. Due to the relatively shallow sensitivity of ASMs, the sample of LMXBs is biased toward nearby sources that exhibit bright X-ray outbursts.

Prior to 2020, the most sensitive all-sky X-ray imaging survey was carried out in 1990/1991 by \rosat at 0.1--2.4\,keV \citep{Truemper1982, Voges1999}. It cataloged X-ray sources brighter than $\sim10\,{\rm \mu Crab}$, providing the deepest X-ray all-sky reference at the time \citep{Boller2016}. Three decades after \rosat, the dynamic X-ray sky is being surveyed by the eROSITA (0.2--10\,keV; \citealt{Predehl2021}) and the Mikhail Pavlinsky ART-XC (4--30\,keV; \citealt{Pavlinsky2021}) telescopes on board the \textit{Spektrum-Roentgen-Gamma} (\srg) mission \citep{Sunyaev2021}. This planned four-year survey obtaining full-sky images created every six months is a powerful X-ray time domain facility. The first eROSITA All-Sky Survey (eRASS1; 2019 December--2020 June) was sensitive to point sources down to $\sim 0.8\,{\rm \mu Crab}$ \citep{Predehl2021}. 

\subsection{AT2019wey}
On 2020 March 18, \srg discovered a new X-ray ($\sim 1\,$mCrab) transient, SRGA\,J043520.9$+$552226 ($=$SRGE\,J043523.3$+$552234; \citealt{Mereminskiy2020}). It coincided with an optical ($r\sim17.5$) transient, \target, first reported by ATLAS \citep{Tonry2019}. 
This transient, bright at both X-ray and optical wavelengths, and located
at low Galactic latitude ($b = 5.3^\circ$) was not present in the Palomar Observatory Sky Survey or the \rosat catalog. We conducted an extensive follow-up campaign, revealing that \target is a Galactic LMXB with unique properties.

\citet[][hereafter Paper I]{Yao2020} presented X-ray observations of \target from 2019 January to 2020 November, suggesting that \target is a LMXB with a BH or NS accretor. In this work, we present multi-wavelength observations of \target. We conclude that the compact object is probably a BH and the companion star must be of low mass ($<1\,M_\odot$). We therefore call \target a candidate BH LMXB. This class of objects and the classification of their X-ray states is reviewed in \citet{McClintock2006, Remillard2006, Belloni2011, Zhang2013, Tetarenko2016}.

The paper is organized as follows. The association between the optical and X-ray transients is outlined in Section~\ref{sec:discovery}. We present optical and ultraviolet (UV) photometry in Section~\ref{sec:photometry}, 
optical and near-infrared (NIR) spectroscopy in Section~\ref{sec:spectroscopy}, and radio observations in  Section~\ref{sec:radio}. We discuss the nature of the source in Section~\ref{sec:discussion}, and summarize out findings and conclusions in Section \ref{sec:conclusion}.

Throughout this paper, times are reported in UT. 
Optical magnitudes are reported in the AB system.
We adopt the reddening law of \citet{Cardelli1989} with $R_V=3.1$.

\section{Association between the Optical and X-ray Transients} \label{sec:discovery}

On 2019 December 2 05:18:40 (MJD 58819.2213), the Zwicky Transient Facility (ZTF; \citealt{Bellm2019b}; \citealt{Graham2019}) detected AT2019wey at a $g$-band ($\lambda_{\rm eff}=4810\,$\AA) magnitude of $g_{\rm ZTF} = 19.30\pm0.05$.
The last non-detection was obtained by ATLAS at an $o$-band ($\lambda_{\rm eff}=6790\,$\AA) magnitude of $o_{\rm ATLAS} > 18.3$, on 2019 December 1 12:18:30 (MJD 58818.5129).

\begin{figure}[htbp!]
	\centering
	\includegraphics[width=\columnwidth]{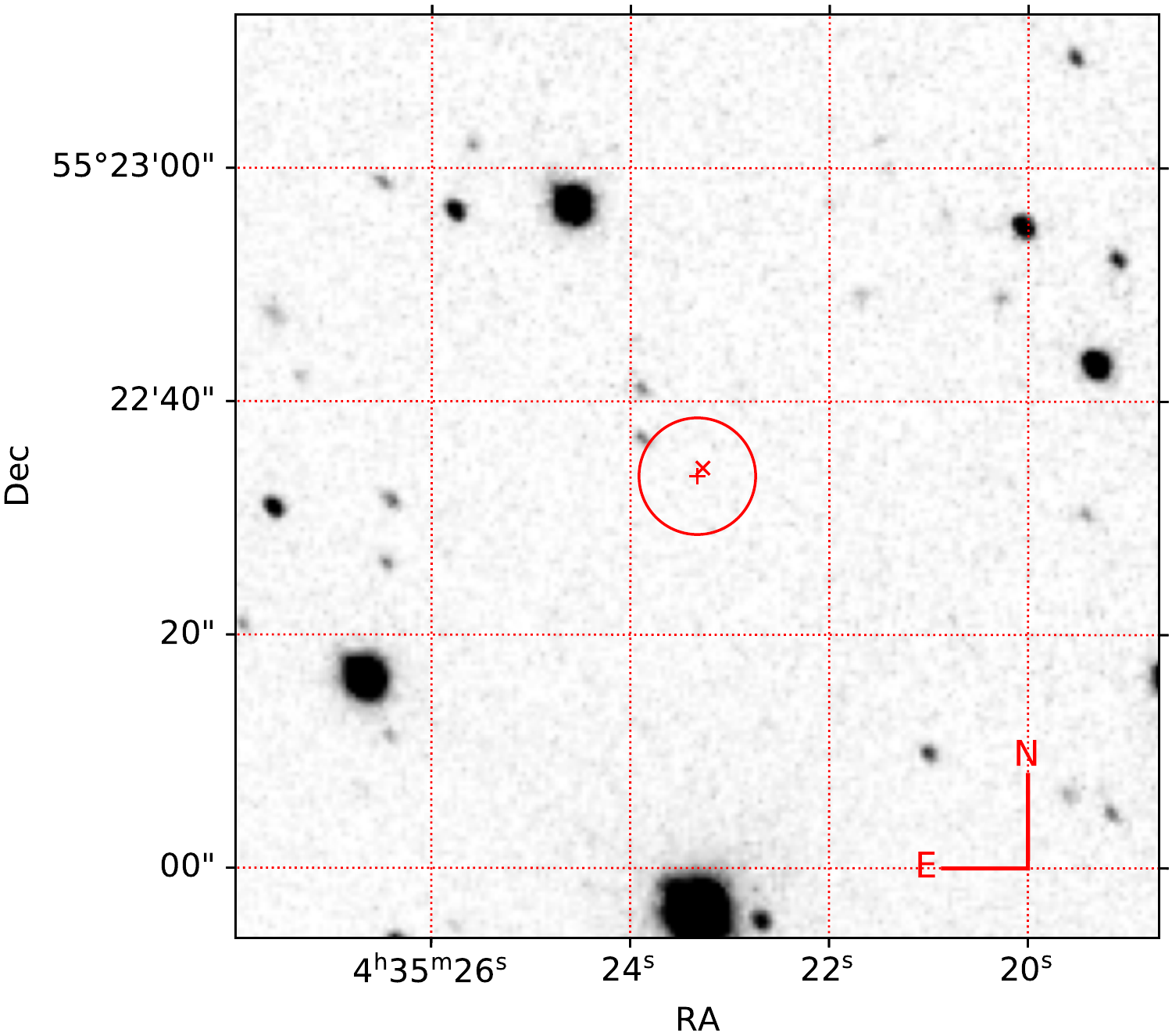}
	\caption{Localization of \target plotted on top of the SDSS $z$-band image. The 
	eROSITA and ZTF position  is shown by ``+'' sign and ``$\times$'', respectively. The circle indicates eROSITA's 68\% error circle radius of 5$^{\prime\prime}$ \citep{Mereminskiy2020}. The ZTF position is R.A=04h35m23.27s, Dec=+55d22m34.3s (J2000).}
	\label{fig:astrometry}
\end{figure}

In Figure~\ref{fig:astrometry} we display the 
X-ray localization on an optical image. The AT2019wey and SRGE\,J043523.3$+$552234 locations are separated by only $0.8^{\prime\prime}$, well within the X-ray error circle radius, thereby confirming the association first suggested by \citet{Mereminskiy2020}. 
The Galactic coordinates of AT2019wey, $l=151.2\degree$ and  $b=5.3\degree$, {\it a priori} favors a Galactic source in the Galactic anti-center direction.

\section{Photometry} \label{sec:photometry}
	
\subsection{ZTF, ATLAS, and \gaia Photometry} \label{subsec:opt_phot}
\begin{figure*}[htbp!]
	\centering
	\includegraphics[width=\textwidth]{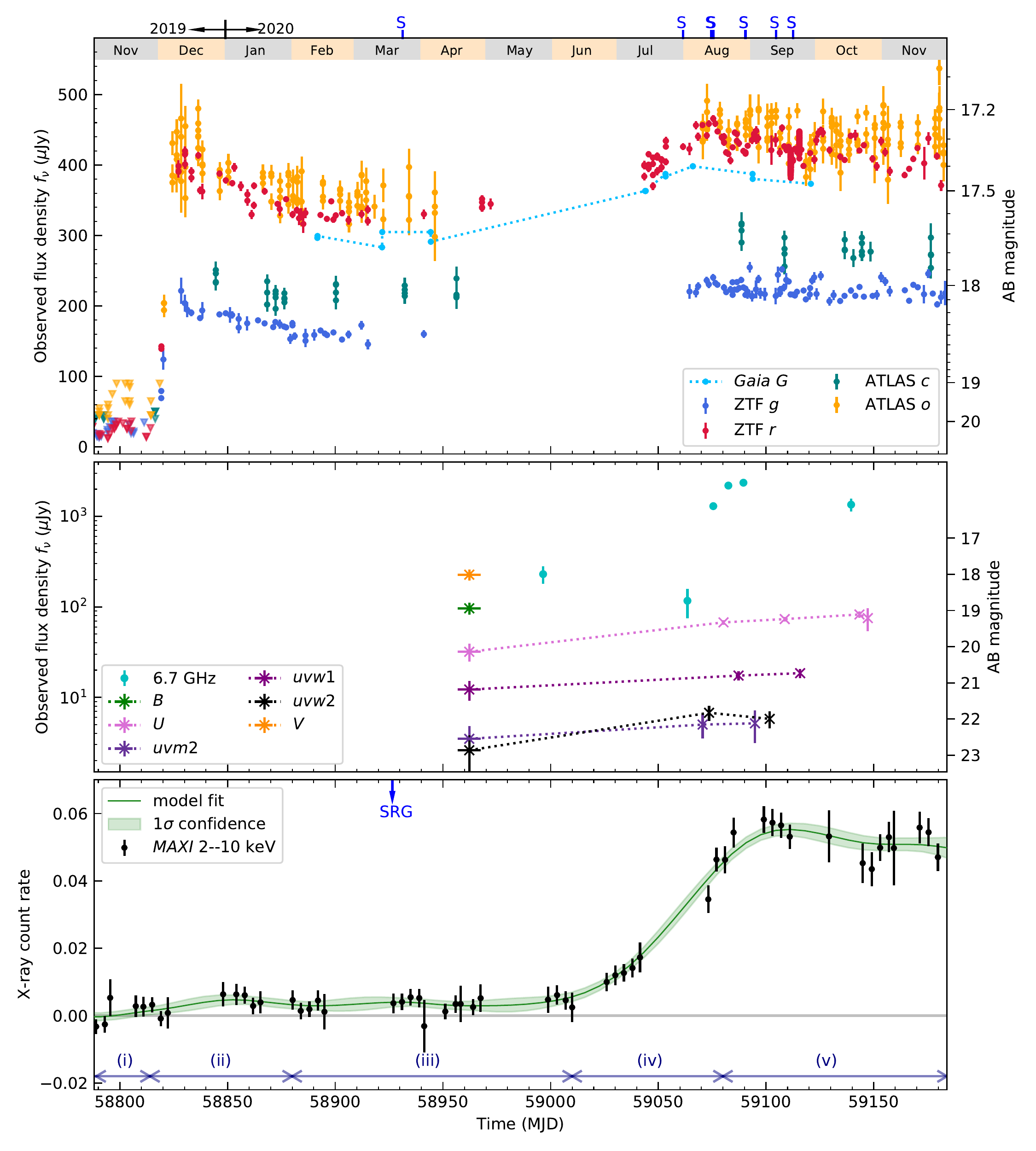}
	\caption{Multi-wavelength light curves of \target. Upper limits are shown as downward triangles. 
	\textit{Upper}: Optical light curves from ZTF, ATLAS, and \gaia (Section~\ref{subsec:opt_phot}).
	Epochs of spectroscopy (Table~\ref{tab:spec}) are marked with the letter S above the upper axis.
	\textit{Middle}: UV (Section~\ref{subsec:uv}) and radio (Section~\ref{sec:radio}) light curves. The 6.7\,GHz flux densities are interpolated from the power-law fits in Table~\ref{tab:vla}. 
	\textit{Bottom}: \maxi 2--10\,keV light curve (Paper I). The green curve is a model fit to the data, generated using a Gaussian process following procedures described in Appendix B.4 of \citet{Yao2020dge}. The epoch of \srg discovery is marked by the blue arrow. The multi-wavelength evolution is divided into five stages (see discussion in Section~\ref{subsec:multi_lc}). 
	\label{fig:multiwave_lc}}
\end{figure*} 


We constructed the optical light curve using the forced-photometry services of ZTF\footnote{\url{https://ztfweb.ipac.caltech.edu/cgi-bin/requestForcedPhotometry.cgi}} \citep{Masci2019} and ATLAS\footnote{\url{https://fallingstar-data.com/forcedphot/}} \citep{Smith2020}.
We obtained \gaia photometry from the \gaia alerts page\footnote{\url{http://gsaweb.ast.cam.ac.uk/alerts/alert/Gaia20aua/}}. See Table~\ref{tab:phot} for the ZTF photometry.

The upper panel of Figure~\ref{fig:multiwave_lc} shows the ZTF, ATLAS, and \gaia light curves of \target. Over the first two weeks, the light curve rose to $r_{\rm ZTF}=17.3$\,mag. After that, the light curve displayed small amplitude ($\lesssim0.3$\,mag) variability for more than 300\,days. The lack of photometry between ${\rm MJD}\sim58980$ to ${\rm MJD}\sim 59040$ is due to the source being in the day sky. On September 9 and 13 we undertook continuous observations as part of the ZTF ``deep drilling'' program \citep{Kupfer2021}. On each day, $\approx 130$ $r$-band exposure frames were obtained.

\subsection{CHIMERA Photometry}
\begin{figure}[htbp!]
	\centering
	\includegraphics[width=\columnwidth]{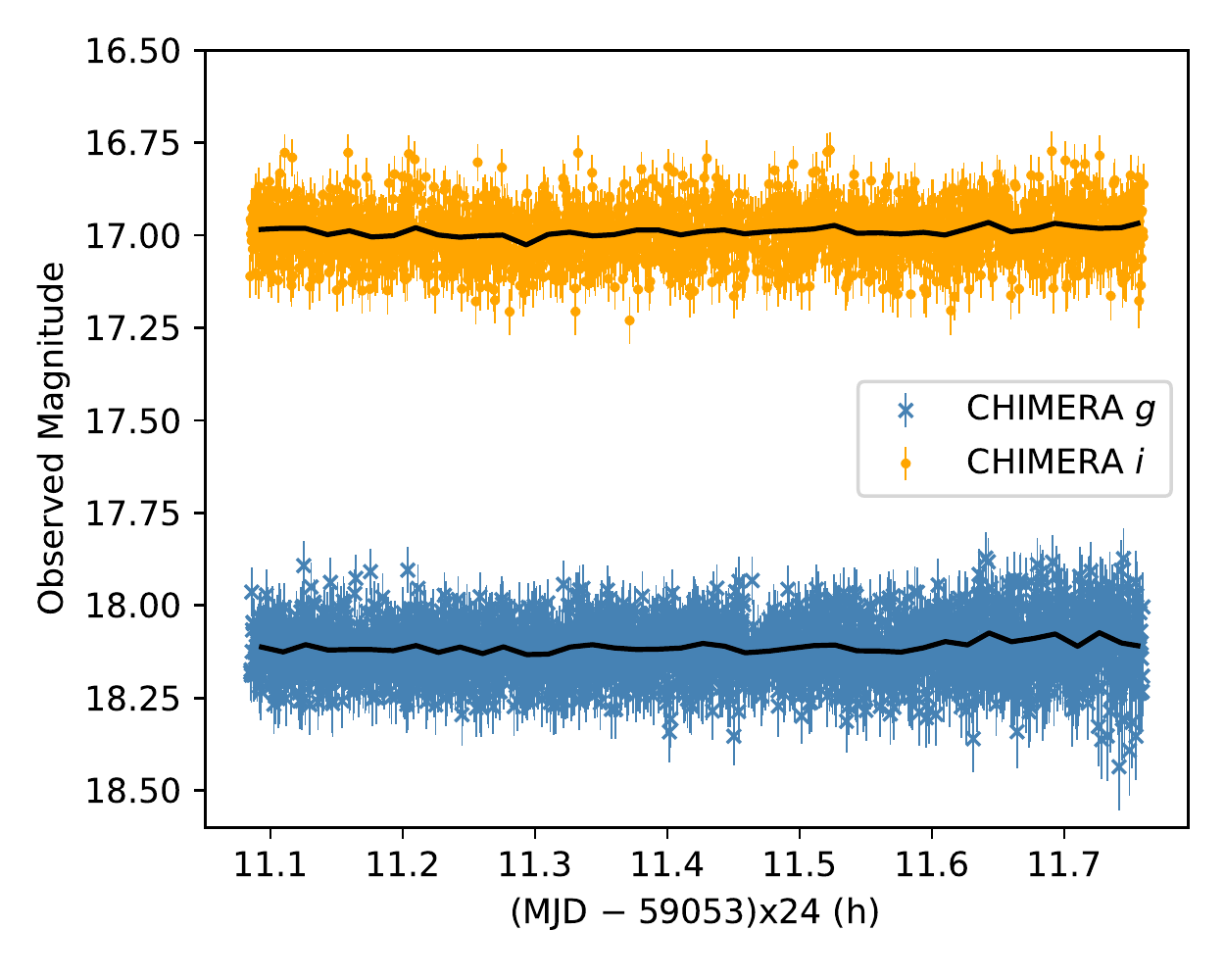}
	\caption{CHIMERA data of \target. The black lines show light curves averaged to 1\,min. The median magnitudes are $i=16.99\pm 0.07$  and $g=18.12\pm 0.08$. The $g$-band rms increased toward the end of the observation due to the onset of twilight.
	\label{fig:chimera_lc}}
\end{figure}

On 2020 July 23 (MJD 59053), we obtained high-speed photometry in the SDSS $g$ and $i$ band using the Caltech HIgh-speed Multi-color camERA (CHIMERA; \citealt{Harding2016}) on the 200-inch Hale telescope of the  Palomar Observatory. We operated the detectors using the 1\,MHz conventional amplifier in frame-transfer mode with a frame exposure time of 1\,s, and obtained 3300 frames in each filter. We reduced the data with a custom pipeline\footnote{\url{https://github.com/mcoughlin/kp84}}. 
Figure~\ref{fig:chimera_lc} shows the CHIMERA light curve. \target appears to exhibit intra-night variability of $\sim0.1$\,mag.

\subsection{Period Search} \label{subsec:psearch}

\begin{figure}[htbp!]
    \centering
    \includegraphics[width = \columnwidth]{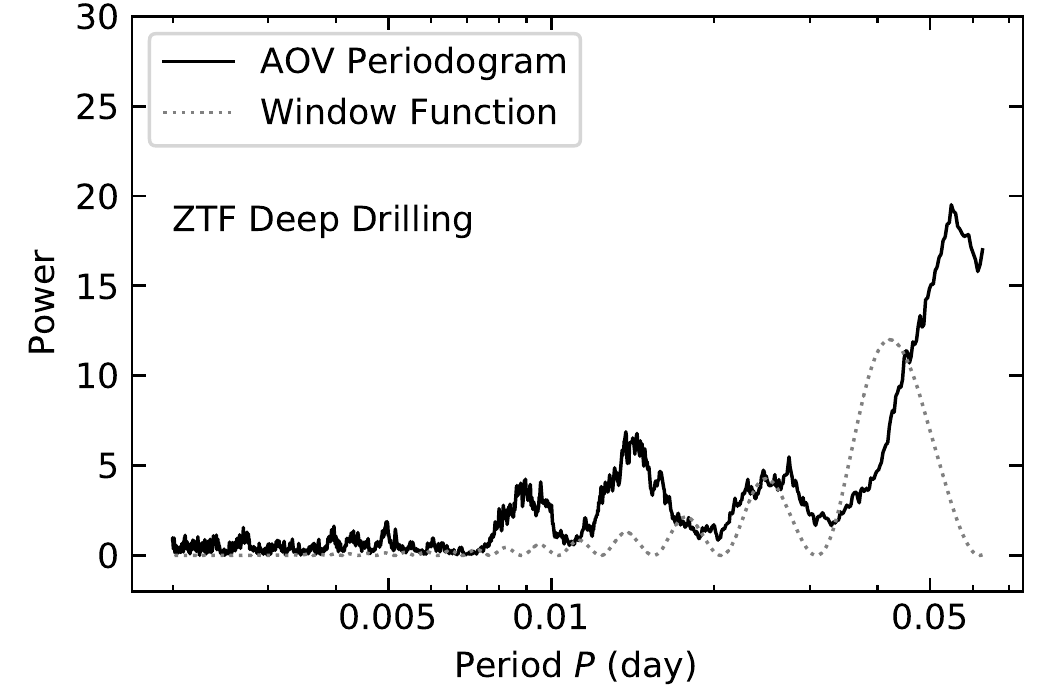}
    \includegraphics[width = \columnwidth]{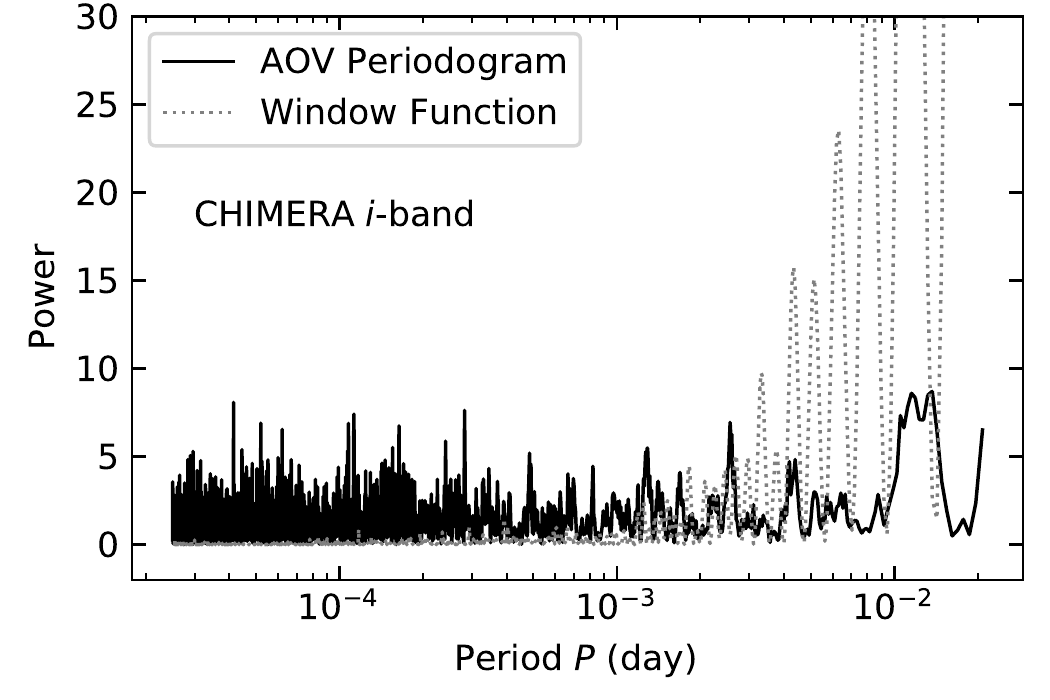}
    \includegraphics[width = \columnwidth]{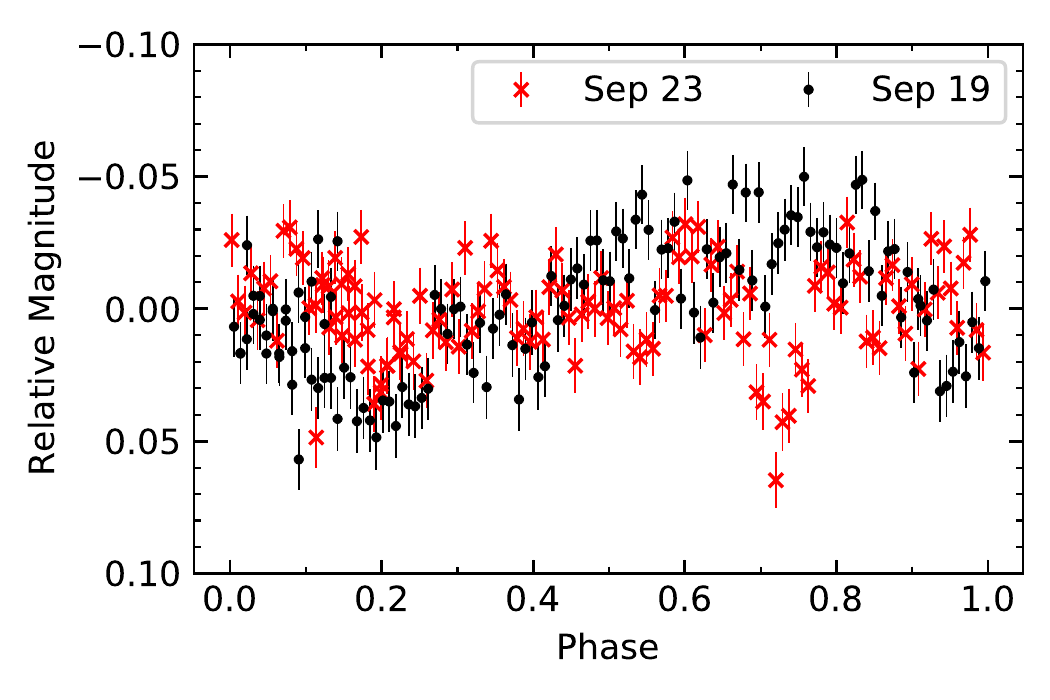}
    \caption{\textit{Upper}: The periodogram and window function for the ZTF deep drilling dataset.
    \textit{Middle}: The periodogram and window function for CHIMERA $i$ band. Note that the periodogram for the $g$-band data is similar to that of the $i$-band.
    \textit{Bottom}: The ZTF deep-drilling light curve, relative to the median, folded on a period of 0.055\,d. \label{fig:periodogram}}
\end{figure}

We ran a periodicity search on the CHIMERA and the ZTF deep drilling datasets using the analysis of variance (AOV) method \citep{Schwarzenberg-Czerny1998}\footnote{We used the \texttt{python} script provided by \url{https://users.camk.edu.pl/alex/\#software}}. We used a frequency grid from 16\,d$^{-1}$ to 500\,d$^{-1}$ for the ZTF data, and a frequency grid from 48\,d$^{-1}$ to 40,000\,d$^{-1}$ for the CHIMERA data. To see how the observational cadence affects the periodogram, we used the Lomb-Scargle algorithm (see a recent review by \citealt{VanderPlas2018}) to compute the window function.

We define ``significance'' of a period as the maximum value in the periodogram divided by the standard deviation of values across the full periodogram.
A possible period at 0.055\,d (1.3\,hr) at a significance of 9.2 can be seen in the ZTF periodogram (upper panel of Figure~\ref{fig:periodogram}). We note that the 1.3\,hr peak is mainly caused by the sinusoidal-like structure observed on September 19, not the dip-like structure observed on September 23. Since the data on September 19 and 23 do not follow the same trend as a function of phase (see lower panel of Figure~\ref{fig:periodogram}), we consider the possible period at 1.3\,hr to be spurious. No period above 8$\sigma$ can be identified from the CHIMERA periodogram. 

\subsection{UV Photometry} \label{subsec:uv}

We obtained UV observations of AT2019wey with 
the Ultra-Violet/Optical Telescope (UVOT; \citealt{Roming2005}) on board the \textit{Neil Gehrels Swift Observatory} \citep{Gehrels2004} from 2020 April to September. The UVOT data were processed using \texttt{HEASoft}. We extracted the photometry with \texttt{uvotsource} using a $3^{\prime\prime}$ circular aperture. Background counts were estimated in a $10^{\prime\prime}$ source-free circular aperture. 
\target was only marginally detected in April. Therefore, for the April datasets, we undertook photometry on co-added images.

In October 2020, we obtained $U$-band photometry using the Spectral Energy Distribution Machine (SEDM, \citealt{Blagorodnova2018}, \citealt{Rigault2019}) on the robotic Palomar 60-inch telescope (P60, \citealt{Cenko2006}). Data reduction was performed using the \texttt{FPipe} pipeline \citep{Fremling2016}.
The UVOT and SEDM photometry are presented in Table~\ref{tab:uvot} and is shown in the middle panel of Figure~\ref{fig:multiwave_lc}.

\section{Optical and NIR Spectroscopy} \label{sec:spectroscopy}

\begin{figure*}[htbp!]
	\centering
	\includegraphics[width=\textwidth]{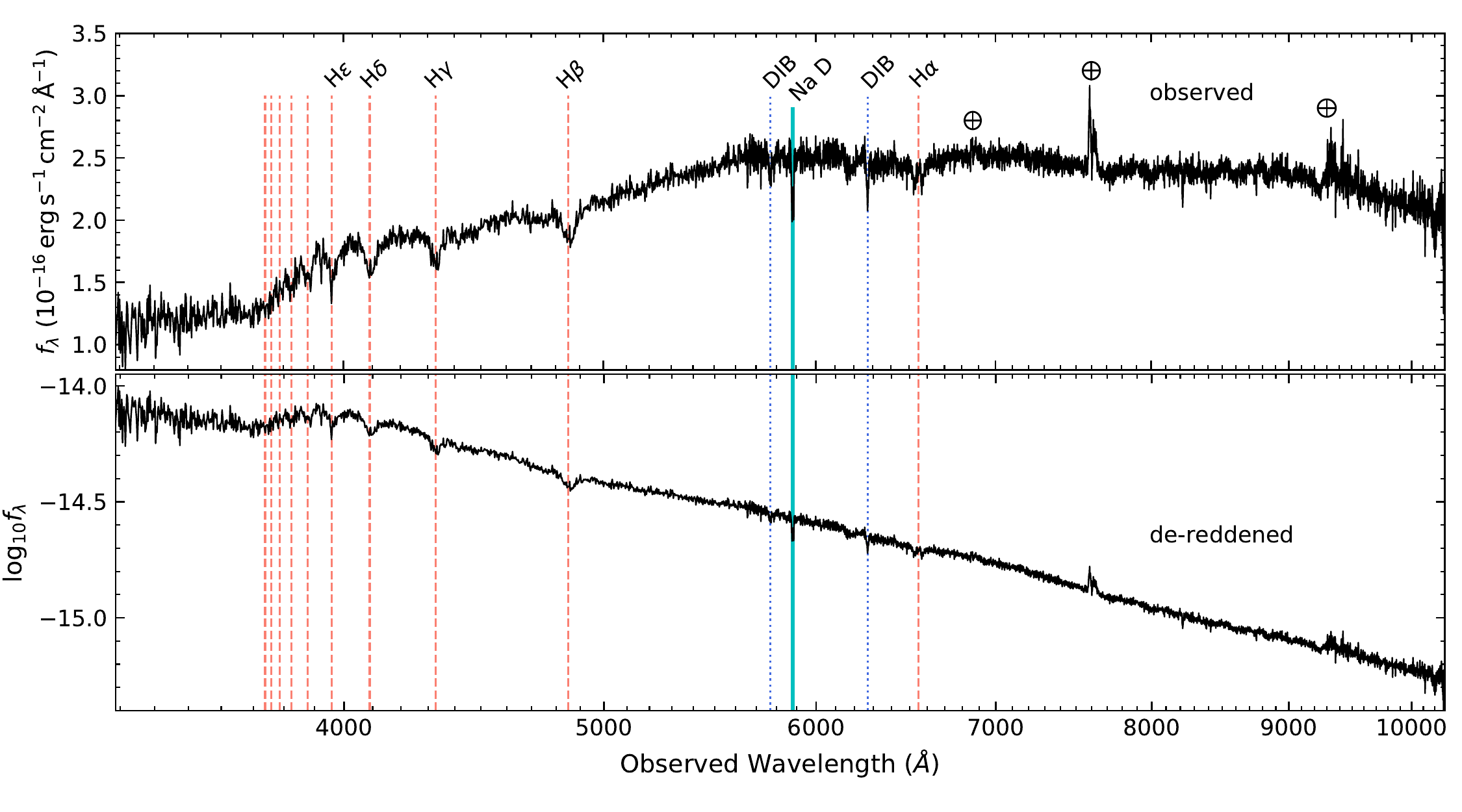}
	\caption{LRIS spectrum of AT2019wey obtained on 2020 March 18. \textit{Upper}: Observed spectrum. \textit{Bottom}: Extinction-corrected spectrum using $E(B-V)=0.9$. Rest (air) wavelength of atomic transitions is marked with vertical lines. \label{fig:spec}}
\end{figure*}

\begin{figure*}[htbp!]
	\centering
	\includegraphics[width=\textwidth]{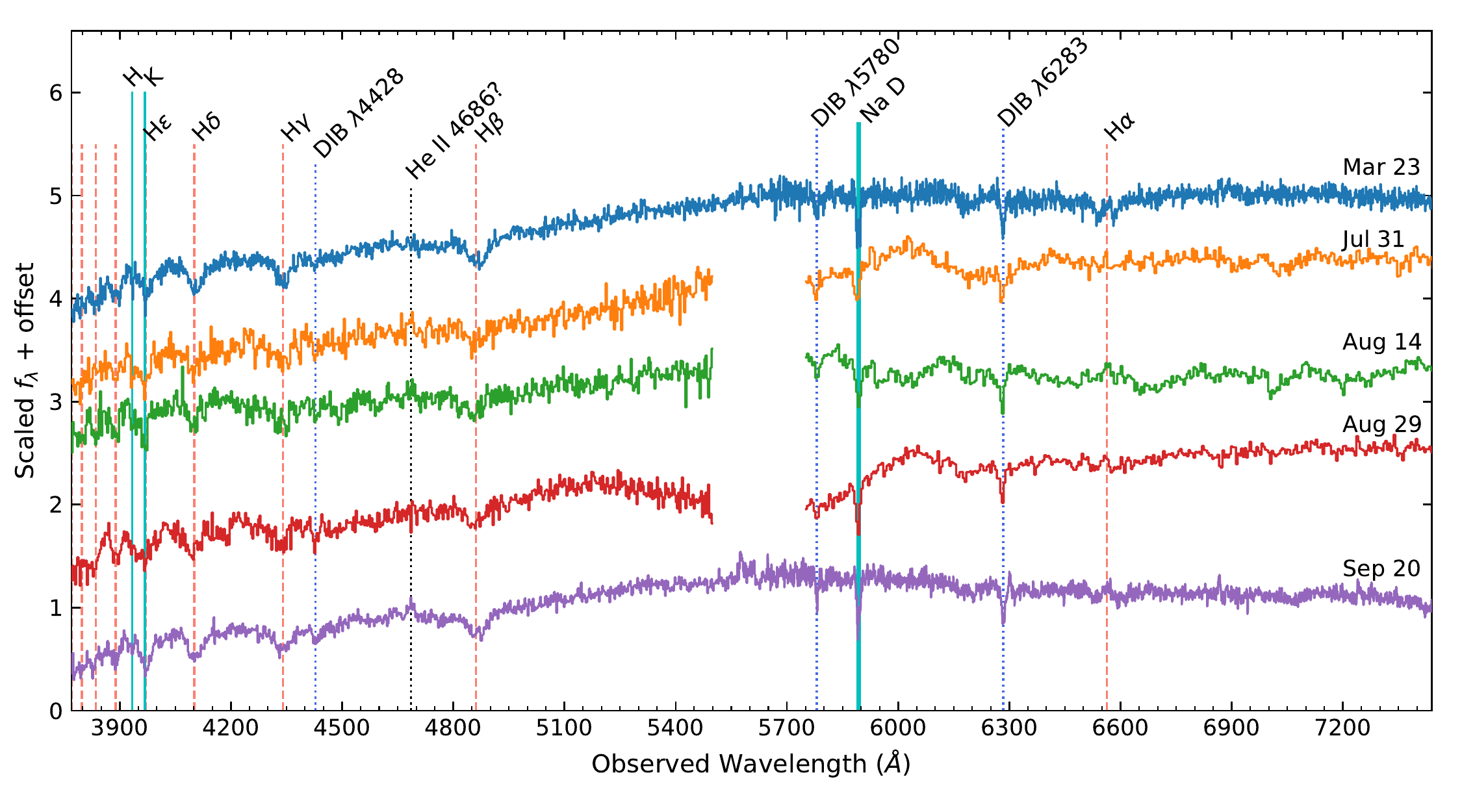}
	\caption{Low-resolution optical spectra of \target (Table~\ref{tab:spec}). We note that during our DBSP observations, the blue end of the red-side CCD had a malfunction, such that flux in the 5550--5750\,\AA\ range was lost (masked).
	\label{fig:spec_bluered}}
\end{figure*}

\begin{deluxetable}{ccccc}[htbp!]
	\tablecaption{Log of AT2019wey spectroscopy. \label{tab:spec}}
	\tablehead{
		\colhead{Date}   
		& \colhead{Telescope/}
		& \colhead{Range}
		& \colhead{Exp.} 
		& \colhead{ Air }\\
		\colhead{in 2020}   
		& \colhead{Instrument}
		&\colhead{(\AA)}
		& \colhead{ (s) }
		& \colhead{ Mass}
	}
	\startdata
	Mar 23 & Keck-I/LRIS & 3200--10250 &  300 & 2.22   \\
	Jul 31 & P200/DBSP & 3410--5550, 5750--9995 & 600 & 1.38  \\
	Aug 13 &Keck-II/NIRES & 9400--24650 & 360 & 1.38  \\
	Aug 14  & P200/DBSP & 3410--5550, 5750--9995 & 600 & 1.34  \\
	Aug 29 &  P200/DBSP &3410--5550, 5750--9995  & 600$\times$2 & 1.40  \\
	Sep 12 & Keck-II/ESI & 3950--10200 &1800 & 1.32 \\
	Sep 20 & Keck-I/LRIS & 3200--10250 & 300$\times$2 & 1.28   \\
	\enddata 
	\tablecomments{All spectra have been uploaded to the TNS page of this source (\url{https://www.wis-tns.org/object/2019wey}). Multiple exposures were obtained on 2020 August 29 and September 20. Since no significant variability was observed on the timescale of 5--10\,min, summed spectra were produced for the two dates.} 
\end{deluxetable}

A log of our spectroscopic observations is given in Table \ref{tab:spec}. The instrumental and observational details can be found in Appendix~\ref{sec:obsinfo}. 

\subsection{Optical Spectroscopy} \label{subsec:opt_spec}

We identify the following features at redshift $z=0$ in all of our spectra: Balmer absorption lines, \ion{Ca}{II} H and K lines, the \ion{Na}{I} D doublet, diffuse interstellar band (DIB) $\lambda 5780$, $\lambda 6283$ absorption features, and the Balmer jump (Figure~\ref{fig:spec}, \ref{fig:spec_bluered}).
\ion{He}{II} $\lambda4686$ emission seems to be detected in the spectra obtained on July 31, August 14, and September 20. We conclude that AT2019wey is a transient of Galactic stellar origin. 

\begin{figure*}[htbp!]
	\centering
	\includegraphics[width=\textwidth]{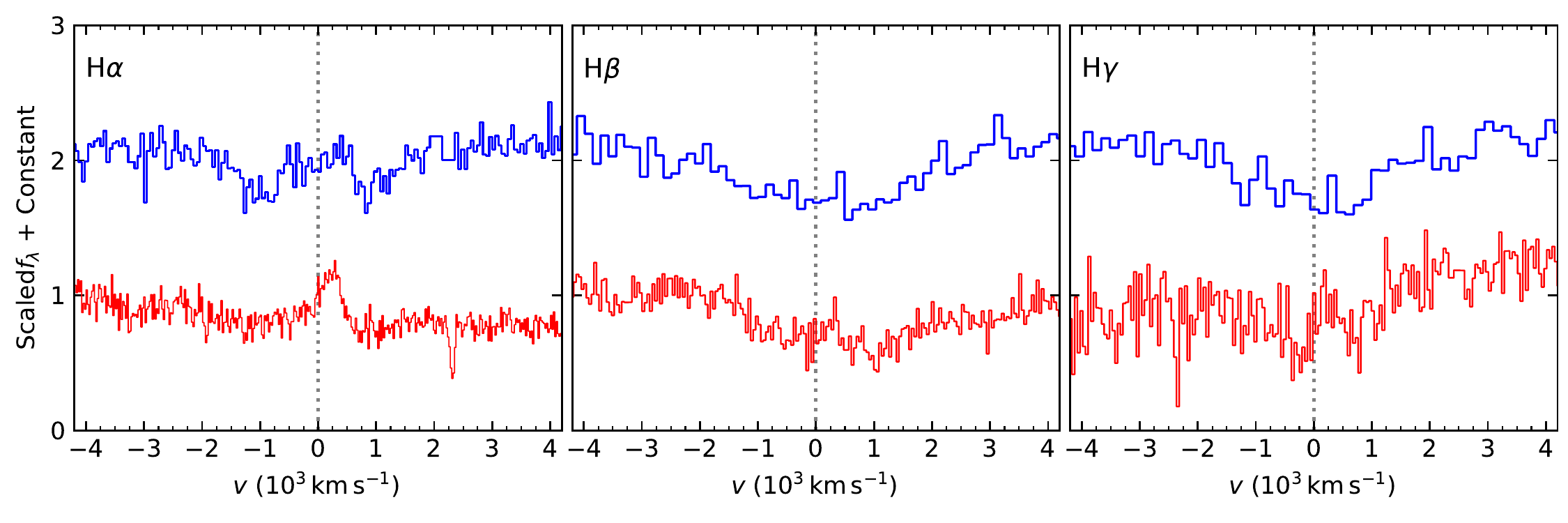}
	\caption{Velocity of the Balmer lines. The 2020 March 23 LRIS spectrum is shown on the top (in blue) and the 2020 September 12 ESI spectrum is shown on the bottom (in red).\label{fig:velocity}}
\end{figure*}

From March to September, the hydrogen profile clearly changed (Figure~\ref{fig:spec_bluered}). 
Figure~\ref{fig:velocity} presents the velocity profiles of Balmer lines in the March 23 and the September 12 spectra. 
On March 23, we observed a relatively narrower (${\rm FWHM}\sim 1200\,{\rm km\, s^{-1}}$) emission component in the middle of a rotationally broadened (${\rm FWHM}\sim 2700\,{\rm km\, s^{-1}}$) shallow absorption trough. At the same epoch, we also observed broad H$\beta$ and H$\gamma$ absorption features with ${\rm FWHM}\sim 2000$--$3000\,{\rm km\, s^{-1}}$. There was a marginal detection of narrow emission cores redshifted by $\sim 300$--$400\,{\rm km\, s^{-1}}$ from the line center of the absorption troughs. 
On September 12, we observed flat-topped H$\alpha$ in emission ($\sim 400\,{\rm km\, s^{-1}}$), while the H$\beta$ and H$\gamma$ profiles were similar to the H$\alpha$ profile on March 23. 
The variable Balmer features are discussed further in Section~\ref{subsec:balmer}.

The reddening of AT2019wey can be constrained to $0.8<E(B-V)<1.2$ (Appendix~\ref{subsec:extinction}) using the equivalent width ($EW$) of the interstellar absorption lines. We find a lower limit to the distance of AT2019wey of $D>1$\,kpc using the velocities of the \ion{Na}{I} doublet in the ESI spectrum (Appendix~\ref{subsec:distance}). In addition, since \target is in the Galactic anti-center direction, the distance to \target is likely less than $\sim$10\,kpc.
Taken together, we conclude that the distance of \target is between $\sim$1\,kpc and $\sim$10\,kpc.

\subsection{NIR Spectroscopy}
\begin{figure*}[htbp!]
	\centering
	\includegraphics[width=\textwidth]{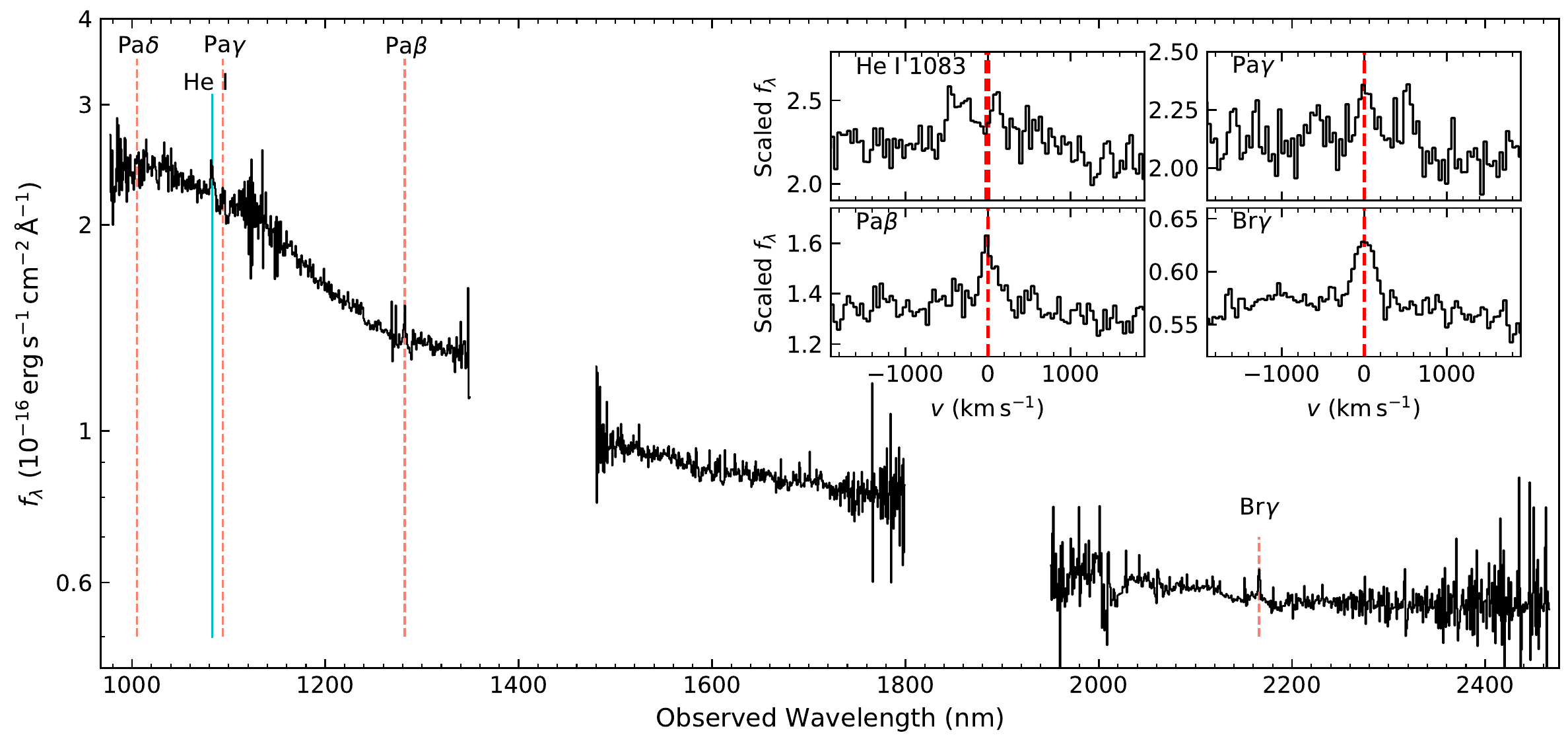}
	\caption{NIRES spectrum of AT2019wey. The insets show the zoom-in on emission lines in velocity 
	space.  
	\label{fig:spec_nir}}
\end{figure*}

The NIR spectrum of \target is shown in Figure~\ref{fig:spec_nir}. Hydrogen emission lines of Pa$\gamma$, Pa$\beta$, and Br$\gamma$ are clearly distinguished. We tentatively attribute the emission lines around 1083\,nm to double-peaked \ion{He}{I}. No absorption lines or molecular bands from the secondary star can be identified. With a FWHM of $\approx 200$--$300\,{\rm km\,s^{-1}}$, the velocities of NIR emission features are much narrower than the H$\alpha$ line, hinting at different formation locations in the accretion disk.

\section{Radio Observations} \label{sec:radio}

\begin{deluxetable}{crrr}[htbp!]
	\tablecaption{Radio observations of \target.\label{tab:vla}}
	\tablehead{
		\colhead{Date}   
		& \colhead{$\nu_0$ (GHz)} 
		& \colhead{$f_\nu$ ($\mu$Jy)}
		& \colhead{$\alpha$}
	}
	\startdata
\multirow{3}{*}{2020-05-27} & 5.0 & $197 \pm 20$ & \multirow{3}{*}{$0.51 \pm 0.69$}\\
&  6.0 & $220 \pm 22$  \\
&  7.0 & $234 \pm 23$  \\
\hline
\multirow{3}{*}{2020-08-02} & 2.5 & $218 \pm 49$  & \multirow{3}{*}{$-0.82 \pm 0.23$}\\
&  3.5 & $205 \pm 16$  \\
&  10.0 & $82 \pm 11$ \\
\hline
\multirow{7}{*}{2020-08-14} & 1.5 & $1023 \pm 75$  & \multirow{7}{*}{$0.23 \pm 0.02$}\\
&  2.5 & $998 \pm 59$  \\
& 3.5 & $1077 \pm 18$  \\
 & 8.5 & $1420 \pm 12$  \\
 & 9.5 & $1399 \pm 11$ \\
 & 10.5 & $1447 \pm 13$  \\
 & 11.5 & $1431 \pm 13$  \\
\hline
\multirow{7}{*}{2020-08-21} & 1.5 & $1676 \pm 102$ & \multirow{7}{*}{$0.19 \pm 0.01$} \\
& 2.5 & $1767 \pm 51$  \\
& 3.5 & $1923 \pm 18$  \\
& 8.5 & $2340 \pm 18$  \\
&  9.5 & $2393 \pm 18$  \\
&  10.5 & $2376 \pm 18$  \\
&  11.5 & $2353 \pm 19$ \\
\hline
\multirow{7}{*}{2020-08-28} & 1.5 & $1846 \pm 128$ & \multirow{7}{*}{$0.20 \pm 0.01$} \\
& 2.5 & $1891 \pm 34$  \\
& 3.5 & $2048 \pm 15$  \\
& 8.5 & $2529 \pm 11$  \\
& 9.5 & $2542 \pm 16$  \\
& 10.5 & $2536 \pm 18$ \\
& 11.5 & $2511 \pm 20$  \\
\hline
2020-10-17  & 6.7 & $1350 \pm 220$ & ---\\
\hline
\multirow{7}{*}{2021-02-17} & 1.5 & $1565 \pm 44$ & \multirow{7}{*}{$0.01 \pm 0.01$}\\
& 2.5 & $1394 \pm 16$  \\
& 3.5 & $1435 \pm 10$  \\
& 8.5 & $1658 \pm 12$  \\
& 9.5 & $1553 \pm 13$  \\
& 10.5 & $1407 \pm 11$  \\
& 11.5 & $1295 \pm 11$  \\
\enddata
\tablecomments{$\nu_0$ is central frequency. The spectral index $\alpha$ ($f_\nu \propto \nu^{\alpha}$) is fitted using the Markov chain Monte Carlo (MCMC) approach with \texttt{emcee} \citep{Foreman-Mackey2013}. The uncertainties are calculated using the 90\% quantiles from the MCMC run.}
\end{deluxetable}

\begin{figure}[htbp!]
	\centering
	\includegraphics[width=\columnwidth]{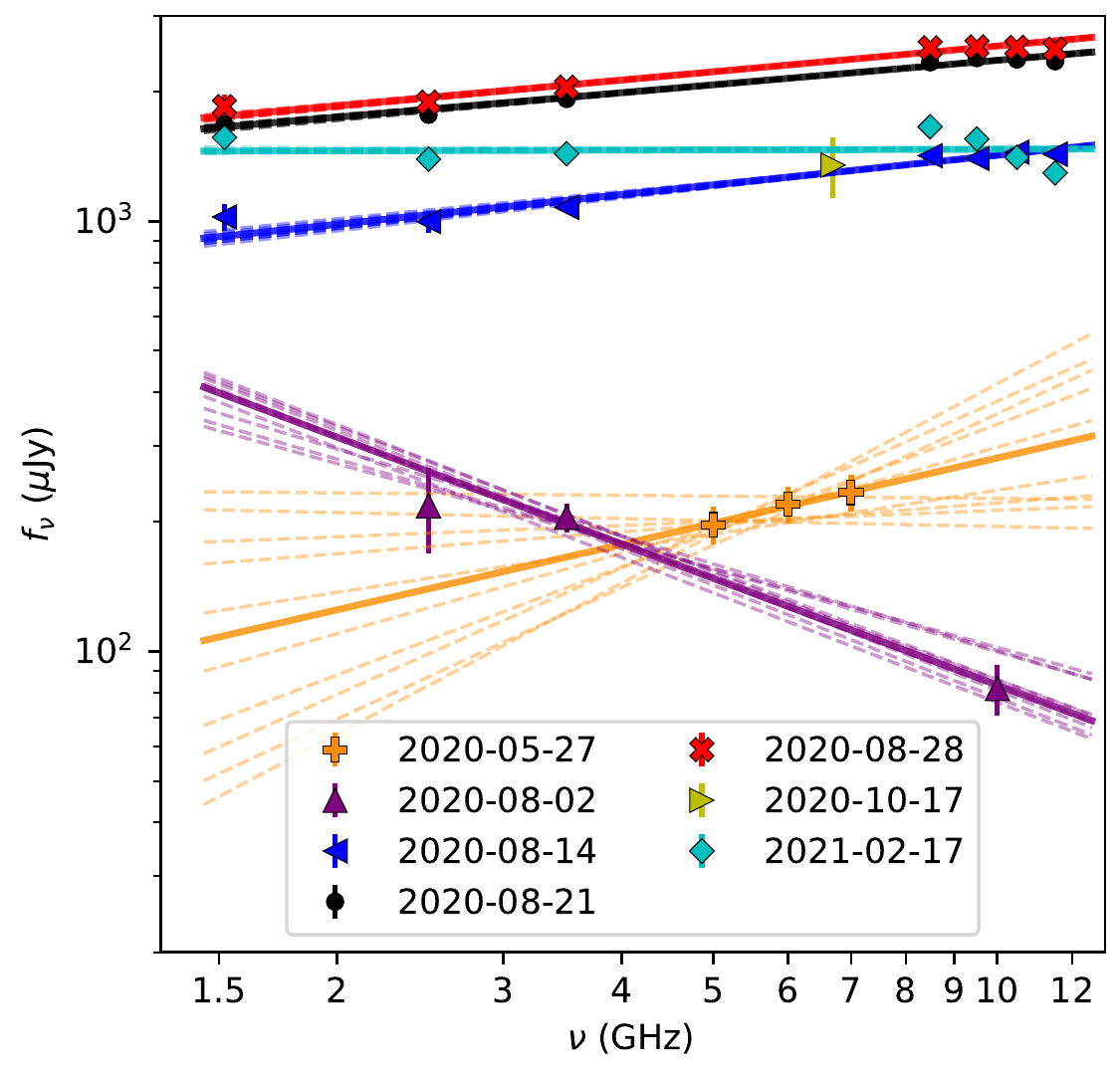}
	\caption{Radio observations of \target. The solid lines are model fits using estimated parameters. Ten random draws from the MCMC posterior are shown with dashed lines. Note that the random draws for the well-constrained models are so well aligned that they cannot be seen.\label{fig:radio_lc}}
\end{figure}

We monitored AT2019wey with the VLA \citep{Perley2011} under the Director's Discretionary Time programs 20A-591 and 20B-397 (PI: Y. Yao). The data were calibrated using the standard VLA Pipeline. We present the flux density of our VLA detections along with the radio detections reported by \citet{Cao2020, Cao2020EVN} in Table~\ref{tab:vla}. We fit a power-law (PL) function ($f_{\nu} \propto \nu^{\alpha}$) to the data; see Figure~\ref{fig:radio_lc} for model fits and Table~\ref{tab:vla} for the value of $\alpha$.

Other than for August 2, the power law fit is flat or slightly inverted ($\alpha\approx0$--0.5). Usually this is attributed to
synchrotron self-absorption and is frequently seen in the low-hard-state (LHS) and hard-intermediate state (HIMS) of X-ray binaries \citep{Fender2001, Fender2004}.
On August 2, however, a ``standard'' spectral index of $\sim -0.8$ was observed. The change of spectral index may indicate the existence of a multi-zone jet. \citet{Yadlapalli2021} reported the detection of a resolved radio source by VLBA in 2020 September, which was interpreted as a steady compact jet.


\section{Discussion} \label{sec:discussion}

The archival (historical) optical data (see Appendix~\ref{subsec:optical_limit}) establish a faint quiescent
counterpart: $r_{\rm SDSS}>22.6$. For $0.8<E(B-V)<1.2$, the 
corresponding extinction is $2.2<A_{r_{\rm SDSS}}<3.3$.
Combined with our distance limit of $D<10$\,kpc, this restricts the
donor star to have an absolute magnitude of $M_R > 4.3$. For a main sequence star, this corresponds to a spectral type later than G2 and a stellar mass $<1\,M_\odot$. For a subgiant star, the stellar mass is even smaller. Therefore, the companion is a low mass ($<1\,M_\odot$) late-type, likely evolved star.
The optical outburst amplitude for \target is $\Delta r > (22.6 - 17.4) = 5.2$\,mag. Using an empirical relation between $\Delta r$ and $P_{\rm orb}$ for short-period LMXBs  \citep{Shahbaz1998}, we find the orbital period, $P_{\rm orb}\lesssim 16$\,hr. 

\subsection{Radio--X-ray Correlation} \label{subsec:LrLx}
\begin{figure}[htbp!]
	\centering
	\includegraphics[width=\columnwidth]{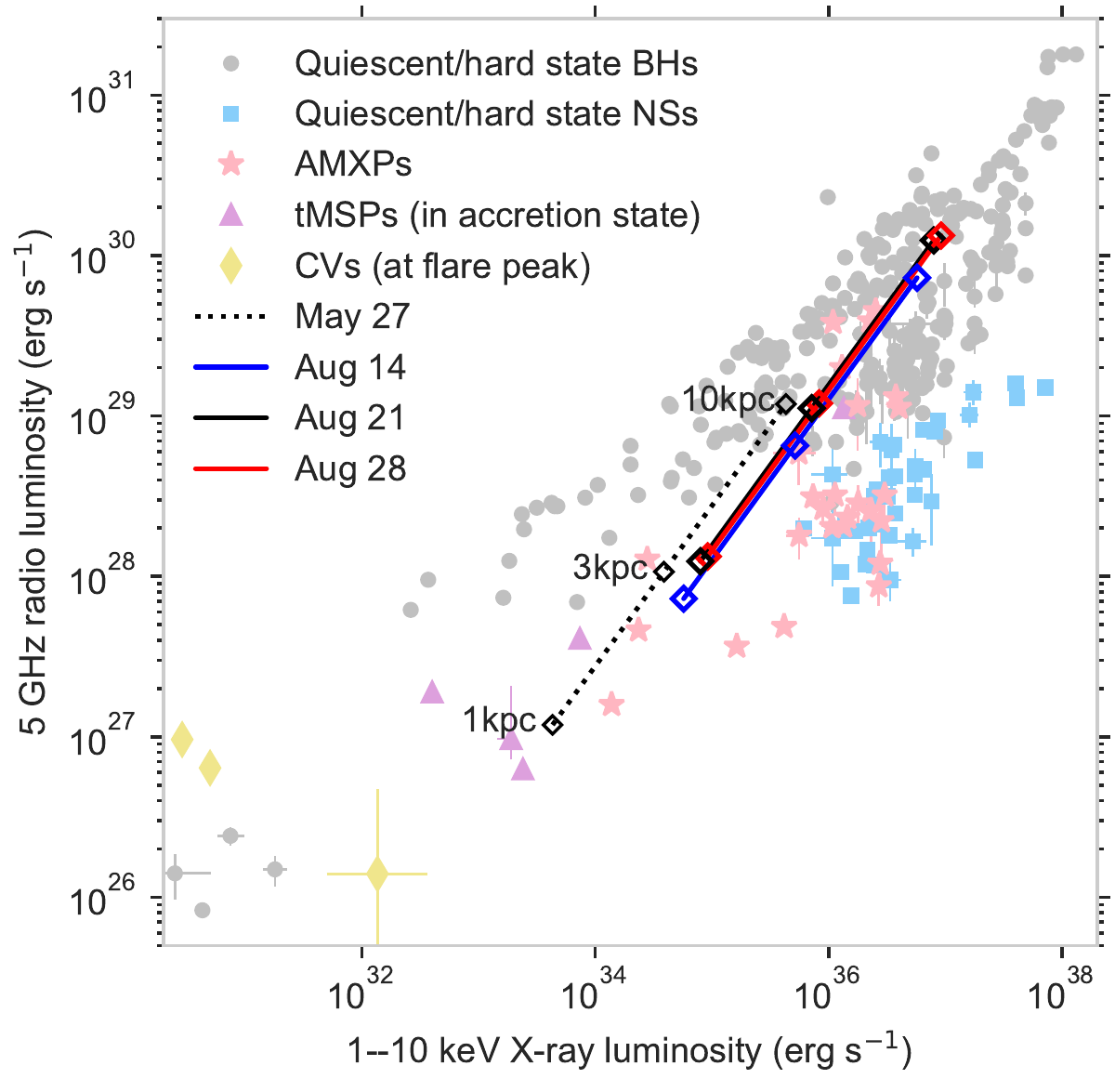}
	\caption{The $L_{\rm radio}$--$L_{\rm X}$ diagram of AT2019wey and various populations of X-ray sources, including quiescent/hard state BHs, NSs, accreting millisecond X-ray pulsars (AMXPs), transitional millisecond pulsars (tMSPs), and cataclysmic variables (CVs) \citep{arash_bahramian_2018_1252036}. We mark the positions of \target at four epochs for possible distances of 1--3--10\,kpc. \label{fig:lrlxplot}}
\end{figure}

Figure~\ref{fig:lrlxplot} shows that on the $L_{\rm radio}$--$L_{\rm X}$ diagram, the position of AT2019wey is above the region occupied by the majority of NS binaries and is closer to BH binaries. Therefore, the bright radio luminosity favors a BH accretor.

\subsection{Multi-wavelength Light Curve}\label{subsec:multi_lc}
We separate the multi-wavelength light curve of \target into five stages (see the bottom panel of Figure~\ref{fig:multiwave_lc}): (i) Before ${\rm MJD}\sim 58814$, the source was in quiescence; (ii) From ${\rm MJD}\sim58814$ to ${\rm MJD}\sim 58880$, the optical light curve exhibited a fast-rise linear-decay outburst, after which it settled onto a $r$-band flux of $f_{\nu{\rm ,}r}\sim  315\,{\rm \mu Jy}$. Around the same time, the X-ray flux rose to $\sim  1$\,mCrab, and stayed in the LHS; (iii) From ${\rm MJD}\sim 58880$ to ${\rm MJD}\sim59010$, the optical and X-ray light curves stayed almost flat; (iv) From ${\rm MJD}\sim 59010$ to ${\rm MJD}\sim59080$, \target exhibited a multi-wavelength brightening, and the X-ray remained in the LHS (Paper I); (v) From ${\rm MJD}\sim59081$ to ${\rm MJD}\sim59180$, the source entered into the HIMS (Paper I). The optical stayed around $f_{\nu{\rm ,}r}\sim  400\,{\rm \mu Jy}$, and X-ray stayed around $\sim  20$\,mCrab (Paper I). 

\subsubsection{UV/optical--X-ray Correlation} \label{subsubsec:LoptLx}
\begin{table*}[htbp!]
    \centering
    \caption{X-ray and optical luminosity of \target at different stages of the multi-wavelength evolution. \label{tab:uvoirx}}
    \begin{tabular}{c|c|c|c}
    \toprule
       Stage   & Band & Luminosity & Comments\\
       \hline\hline
       (iii) &   $r$ \& $g$ & $4.0\times 10^{34}$ \& $6.1\times 10^{34}$ & Averaged between ${\rm MJD}\sim58880$ and ${\rm MJD}\sim59010$ \\
       (iii) &   X-ray & $1.0\times 10^{35}$ & Averaged between ${\rm MJD}\sim58951$ and ${\rm MJD}\sim 58967$\\
    \hline
        (v) &   $r$ \& $g$ & $4.9\times 10^{34}$ \& $8.4\times 10^{34}$  & Averaged between ${\rm MJD}\sim 59080$ and ${\rm MJD}\sim 59153$ \\
       (v) &   X-ray & (1.3--1.7)$\times 10^{36}$ & Range of values from minimum ({\rm MJD}$\sim 59082$) to maximum (${\rm MJD}\sim59112$)  \\
    \bottomrule
    \end{tabular}
    \tablecomments{Luminosity is given in units of $(D/5\,{\rm kpc})^2\,{\rm erg\,s^{-1}}$. X-ray column density corrected luminosity is given in 2--10\,keV, assuming $N_{\rm H}=5\times 10^{21}\,{\rm cm^{-2}}$. Optical luminosity has been corrected for extinction, adopting $E(B-V)=0.9$.}
\end{table*}

\begin{figure}
    \centering
    \includegraphics[width=\columnwidth]{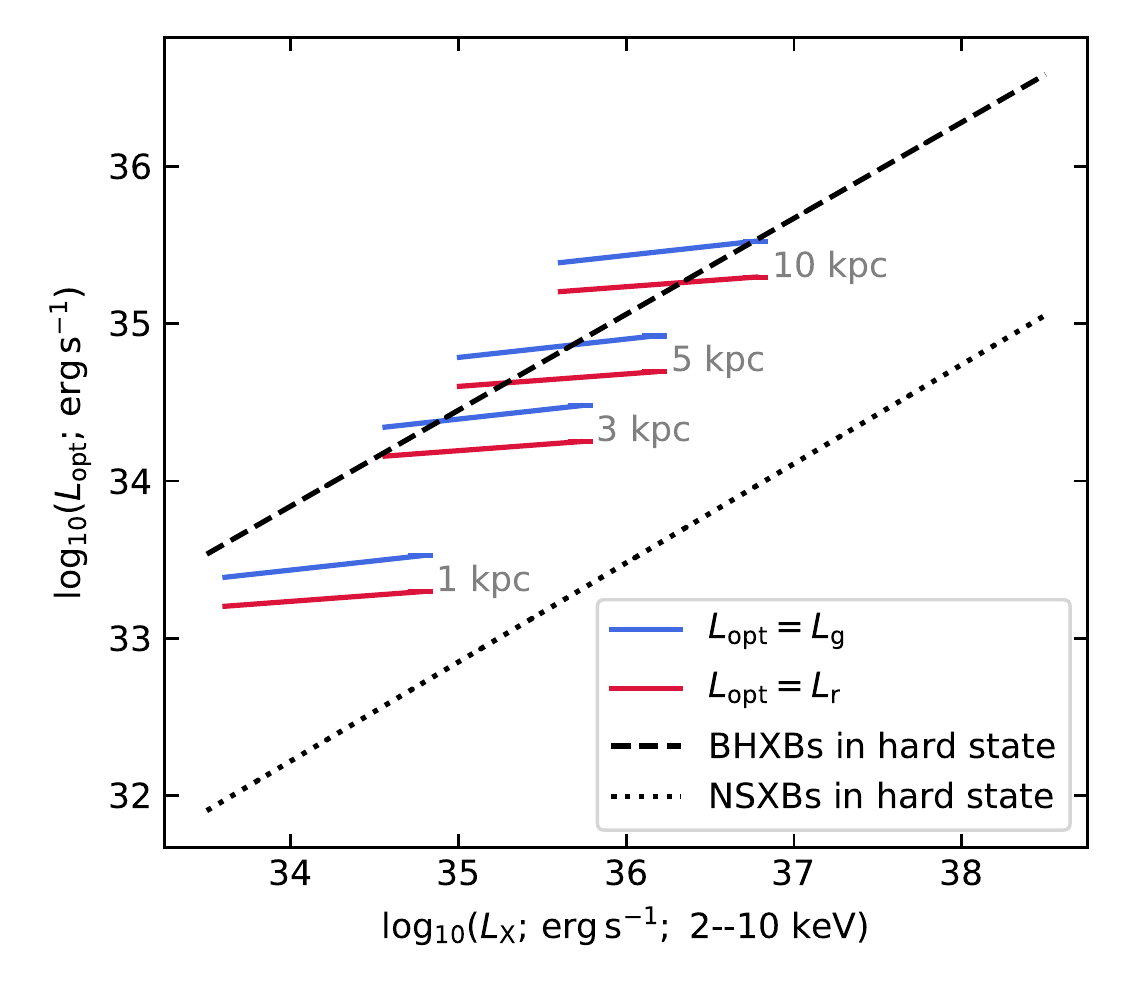}
    \caption{The solid lines demonstrate the correlation between optical ($g$- or $r$-band) and X-ray luminosities of \target from stage (iii) to stage (v), assuming distances at [1, 3, 5, 10]\,kpc. The dashed and dotted lines are best power-law fits to BH X-ray binaries (BHXBs) and NS X-ray binaries (NSXBs) in the hard state, respectively \citep{Russell2006}. \label{fig:LxLopt}}
\end{figure}

During stage (iv), the X-ray and radio fluxes increased by a factor of $\gtrsim 10$ but in the optical/UV the increase was modest, between a factor of 1.3 and 2. During stages (iii) and (v), the source was stable and representative luminosities can be found in Table~\ref{tab:uvoirx}. For these two stages, following
\citet{Russell2006}, we link the UV/optical and X-ray luminosities as
\begin{align}
    L_{\rm UV/opt} = A L_{\rm X}^{\beta},
\end{align}
and find $\beta\sim 0.08$ in $r$ band, $\beta \sim 0.12$ in $g$ band, and $0.12 \lesssim \beta \lesssim 0.34$ in the UV bands. \citet{Russell2006} derived $A = 10^{13.1\pm0.6}$, $\beta = {0.61\pm0.02}$ for a sample of 15 BH LMXBs, and $A = 10^{10.8\pm1.4}$, $\beta={0.63\pm0.04}$ for a sample of 8 NS LMXBs. As can be seen from Figure~\ref{fig:LxLopt},
 over the distance range of $1\lesssim D \lesssim 10$\,kpc, the inferred luminosities of AT2019wey are suggestive of an accreting BH system.

\subsubsection{Possible Mechanisms for the Optical Emission}\label{subsubsec:3process}
In BH LMXBs in the hard state, the optical/UV emission can arise from  (1) X-ray reprocessing in the outer accretion disk; (2) the optically thick jet spectrum extending from centimeter wavelengths; (3) intrinsic thermal emission from the viscously heated outer accretion disk. For processes (1) and (2), the expected slopes are $\beta\sim 0.5$ \citep{vanParadijs1994}, and $\beta\sim 0.7$ \citep{Corbel2003, Russell2006}, respectively. For process (3),
$\beta$ ranges from 0.13 (Rayleigh--Jeans or R--J tail) to 0.33 (between the R--J tail and the Wien cut-off) \citep{Tetarenko2020}. \citet{Russell2006} find $\beta\sim 0.6$ for BH LMXBs, which suggests that process (3) is not dominant. However, the observed $\beta\sim 0.1$ for \target favors process (3).

Curiously, we note that such small values of $\beta$ have been observed in two BH LMXBs with short orbital periods: $\beta\sim 0.2$ \citep{Armas-Padilla2013} in Swift\,J1357.2$-$0933 ($P_{\rm orb} = 2.8$\,h; \citealt{Corral-Santana2013, Mata-Sanchez2015}) and $\beta\sim0.2$ \citep{Chiang2010} in Swift\,J1753.5$-$0127 ($P_{\rm orb} \lesssim 3.2$\,h; \citealt{Zurita2008, Neustroev2014, Shaw2016ii}). Interestingly, the X-rays for these two systems are only observed in the LHS or HIMS, without successful transitions to the high/soft state (HSS) \citep{Armas-Padilla2013, Tetarenko2016}. These similarities might be understood as characteristics of a sub-population of BH LMXBs (\citealt{Shaw2013}, see Section~\ref{sec:conclusion}). 

\begin{figure*}[htbp!]
	\centering
	\includegraphics[width=\textwidth]{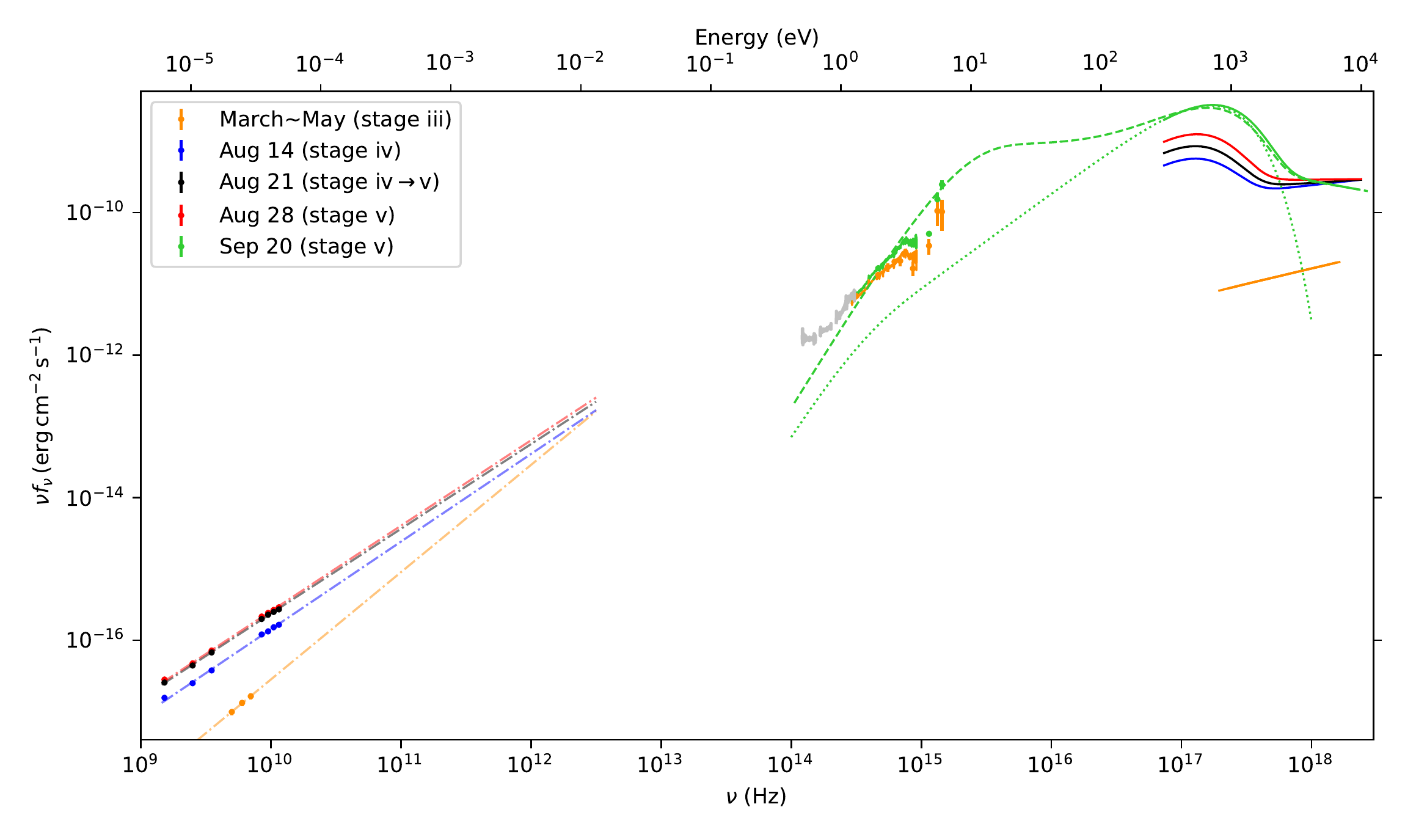}
	\caption{Multi-wavelength SED of AT2019wey. 
	In the radio, we show the observed data and power-law fits (Table~\ref{tab:vla}). In UV/optical/NIR, we show the dereddened photometry and spectra assuming $E(B-V)=0.9$. Note that the silver NIR spectrum, the orange optical spectrum, and the green optical spectrum were obtained on August 13, March 20, and September 20, respectively (Table~\ref{tab:spec}). In X-ray, we show the best fits to X-ray data corrected for a fixed column density of $N_{\rm H}=5\times10^{21}\,{\rm cm^{-2}}$ (see Section~\ref{subsubsec:xray_models} and Paper I).
	See definition of different stages in Section~\ref{subsec:multi_lc}.
	The dashed and dotted green lines from optical to X-ray are illustrative models of irradiation and standard disk emission, respectively (see Section~\ref{subsubsec:opt_bright} for details).
	\label{fig:continuum}}
\end{figure*}

\subsection{Multi-wavelength SED}\label{subsec:multi_sed}

The spectral energy distribution (SED) of \target is shown in Figure~\ref{fig:continuum}. 
The X-ray data are presented in Paper I and we briefly summarize the X-ray spectra in Section~\ref{subsubsec:xray_models}. In
Section~\ref{subsubsec:radio_models}, based on radio data, we conclude that jet emission is unlikely to be the dominant mechanism in the optical. In Section~\ref{subsubsec:opt_dim}, we show that the UV/optical emission during stage (iii) originates from the intrinsic emission of a truncated accretion disk. In Section~\ref{subsubsec:opt_bright}, we show that the UV/optical emission during stage (v) arises from X-ray reprocessing.

\subsubsection{The X-ray SED}\label{subsubsec:xray_models}
Briefly speaking, the X-ray spectrum observed in stage (iii) can be described by an absorbed power-law with photon index $\Gamma=1.8$. In stages (iv) and (v), the X-ray spectrum can be fitted with a combination of disk-blackbody (\texttt{diskbb}, \citealt{SS73, Mitsuda1984}) and power-law components (Paper I).
On August 14, 21, and 28, the fitted models have $\Gamma \sim 1.9$ and inner disk temperature $T_{\rm disk}\sim 0.21\,{\rm keV}\sim 2.4\times 10^{6}$\,K. The inner disk radius is
\begin{align}
R_{\rm in} \sim (360 \textrm{--} 470) \left( \frac{{\rm cos}i}{1}\right)^{-1/2} \left(\frac{D}{5\,{\rm kpc}} \right) \, {\rm km} . \label{eq:Rin}
\end{align}
On September 20, the soft X-ray flux reached a local maximum in the HIMS, where the PL softened to $\Gamma = 2.3$ and the inner disk temperature increased to $T_{\rm disk}\sim 0.29\,{\rm keV}\sim 3.4\times 10^{6}$\,K, while the inner disk radius remains at $\sim 400\,{\rm km}$. The fitted $T_{\rm disk}$ and $R_{\rm in}$ are typical for thermal emission from a truncated accretion disk observed in the LHS and HIMS of BH LMXBs \citep{Done2007}. Denoting the innermost stable circular orbit radius as $R_{\rm ISCO} = 6GM/c^2$ and the Schwarzschild radius as $R_{\rm S} = 2GM/c^2$, then $R_{\rm in}\sim 15 R_{\rm S} \sim 5 R_{\rm ISCO}$ for a 10\,$M_\odot$ non-spinning black hole.

\subsubsection{The Radio SED}\label{subsubsec:radio_models}

The dashed-dotted lines shown in Figure~\ref{fig:continuum} are best-fit power-laws for the radio data (Table~\ref{tab:vla}) extrapolated to $3\times 10^{12}$\,Hz. If the spectrum remains optically thick all the way to the optical and near-infrared (OIR) wavelengths, it will over-predict the observed OIR spectrum. Assuming a classical jet spectrum of a broken PL \citep{Blandford1979}, the break frequency must be $\ll 10^{14}$\,Hz. The optically thin jet spectrum may contribute a fraction of NIR emission (grey data in Figure~\ref{fig:continuum}), but is unlikely to dominate in the optical. 

\subsubsection{UV/Optical Emission in the Dim LHS}\label{subsubsec:opt_dim}
\begin{figure}[htbp!]
    \centering
    \includegraphics[width=\columnwidth]{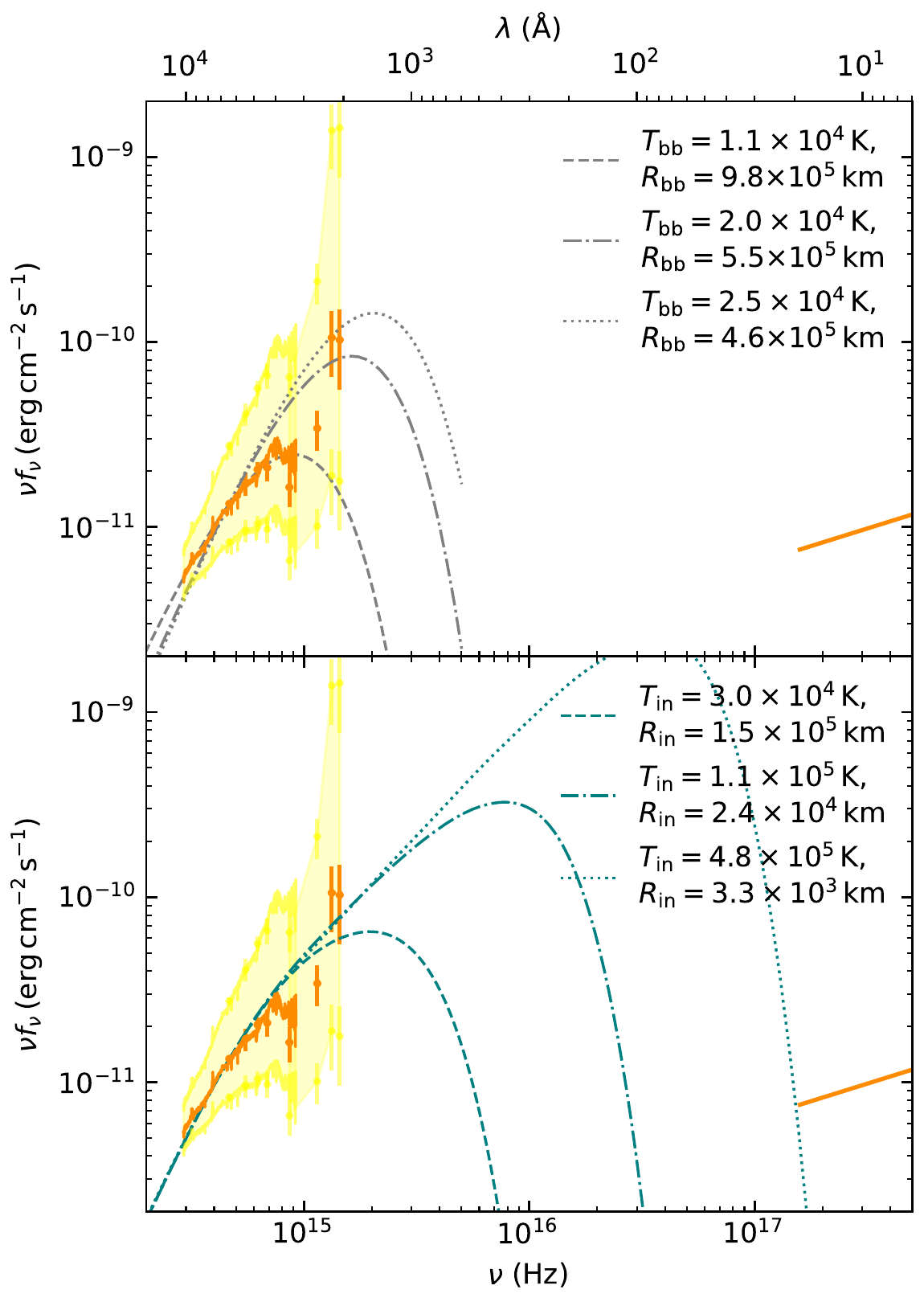}
    \caption{X-ray--UV--optical SED of \target in the dim LHS. Single-temperature blackbody models are shown in the upper panel, while disk-blackbody models are shown in the lower panel. All models are normalized to match the flux in $r$ band. $D=5$\,kpc and ${\rm cos}i=1$ are assumed. The upper and lower bounds of the yellow region are obtained by dereddening the observed data using $E(B-V) = 1.2$ and 0.7, respectively. No detailed model fits are performed due to the uncertainty of $E(B-V)$. \label{fig:opt_sed}}
\end{figure}

In Figure~\ref{fig:opt_sed}, we show the UV/optical data and the best-fit X-ray model in the dim LHS (stage iii) in orange. 
The low level of X-ray flux (compared to that in the UV/optical) suggests that there is not enough X-ray flux to illuminate the outer accretion disk. As a result, the UV/optical probably comes from the intrinsic thermal emission of an accretion disk. 

To obtain a constraint on the outermost annulus of the accretion disk, we compute a set of simple blackbody models (upper panel of Figure~\ref{fig:opt_sed}). 
We adopt the 11,000\,K blackbody as an approximation of the outer disk annulus, and compute a set of \texttt{diskbb} models to obtain a lower limit to the inner disk radius (and an upper limit to the inner disk temperature). The dotted line in the lower panel of Figure~\ref{fig:opt_sed} suggests $T_{\rm in} < 4.8\times 10^{5}$\,K and $R_{\rm in} > 3.3\times 10^{3}\,{\rm km} \sim 38 R_{\rm ISCO} \sim 114 R_{\rm S}$. 

Similar SED shapes have been observed in the LHS of a few BH LMXBs, including XTE\,J1118$+$480 ($R_{\rm in}=300R_{\rm S}$; \citealt{Yuan2005}) and Swift\,J1753.5$-$0127 ($R_{\rm in}>100R_{\rm S}$; \citealt{Froning2014}). The observed SED of \target in the dim LHS fits into the advection-dominated accretion flow (ADAF; \citealt{Narayan1994, Narayan1995oct}) model of a hot accretion flow around a BH, which is predicted at low-accretion rates (see reviews by \citealt{Done2007, Yuan2014, Poutanen2014}).  If so, the X-ray PL comes from a high-temperature flow in the central regions close to the BH, while the UV/optical thermal component comes from a geometrically thin, optically thick accretion disk truncated far from the ISCO \citep{Yuan2014}. 

\subsubsection{UV/Optical Emission in the HIMS}\label{subsubsec:opt_bright}
The dotted green line in Figure~\ref{fig:continuum} shows an extrapolation of the \texttt{diskbb} fit on \nicer data for September 20. It clearly under-predicts the observed UV/optical spectrum, making X-ray reprocessing the most likely origin of the UV/optical emission in the HIMS. 
We therefore attempt to fit the green data by the irradiation model \texttt{diskir} \citep{Gierlinski2008, Gierlinski2009}. 

We set the inner disk temperature of the unilluminated disk and the asymptotic PL photon index to be the same as the best-fit September 20 model (see Section~\ref{subsubsec:xray_models}). The fraction of reprocessed luminosity in the Compton tail ($f_{\rm in}$) is fixed at 0.1. The electron temperature is fixed at 1000\,keV as there is no sign of a high-energy PL cutoff (see Paper I). The dashed green line in Figure~\ref{fig:continuum} is a schematic fit with the following parameters: 
the ratio of luminosity in the Compton tail to that of the unilluminated disk $L_{\rm C}/L_{\rm d}=0.22$, the radius of the Compton illuminated disk $R_{\rm irr} = 1.2 R_{\rm in}$, the fraction of thermalized bolometric flux $f_{\rm out}=0.08$, $R_{\rm out} = 10^{3.55}R_{\rm in}$, and the normalization parameter of the un-illuminated disk (Eq.~\ref{eq:Rin}) $\approx 370$\,km. We conclude that the UV/optical SED in the HIMS is due to reprocessing of the X-ray irradiation.

\subsection{Optical Spectral Lines} \label{subsec:balmer}
The hydrogen lines in \target display both broad absorption and emission components (Section~\ref{subsec:opt_spec}). This behavior is reminiscent of some LMXBs and CVs, where the hydrogen absorption and emission lines are thought to arise from different layers of the viscous accretion disk \citep{Horne1986, LaDous1989, Warner1995}. In a few BH LMXBs, such as
GRO\,J1655$-$40 \citep{Soria2000}, 
GRO\,J0422$+$32 \citep{Callanan1995},
XTE\,J1118$+$480 \citep{Dubus2001, Torres2002},
and Swift\,J1753.5$-$0127 \citep{Rahoui2015}, double-peaked H$\alpha$ was observed. The single-peaked hydrogen line profile of \target is similar to that observed in 
MAXI\,J1836$-$194 \citep{Russell2014}, 
suggesting a binary system viewed close to face-on. This is in agreement with the low inclination ($i\lesssim30^{\circ}$) constraint from modeling the X-ray reflection spectrum (Paper I). 

In Section~\ref{subsubsec:opt_bright} we have shown that in the HIMS, the UV/optical emission comes from the reprocessing of inner disk and coronal emission. Irradiation of the outer disk may form a thin temperature-inversion layer on the disk surface \citep{Tuchman1990}. This naturally explains the enhanced Balmer emission lines observed during stage (iv) and stage (v). 

Most BH LMXBs show strong \ion{He}{II} emission during their outbursts \citep{Zurita2002, Kaur2012, JimenezIbarra2019, Russell2014}. A lack of significant \ion{He}{II} was observed in the optical spectra of AT2019wey. This might also be present in the 2009 outburst of XTE\,J1752$-$223 \citep{Torres2009}, and the 2021 outburst of XTE\,J1859$+$226 (\citealt{Bellm2021}, Bellm et al. in prep). 
We note that the \ion{He}{II} recombination line was also not significantly detected in the outburst spectra of a few CVs \citep{Morales-Rueda2002}. A possible explanation for this is that the number of photons with energies between 54\,eV (the ionization potential of He$^{+}$) and 280\,eV (the ionization potential of the carbon K-edge) is not large enough \citep{Patterson1985}.

\section{Conclusion} \label{sec:conclusion}
We have undertaken a detailed multi-wavelength follow-up of the X-ray transient \target. This study builds upon X-ray observations reported in Paper I, which show that \target is a LMXB with a NS or BH accretor. In this paper, we present the high radio (Section~\ref{subsec:LrLx}) and optical (Section~\ref{subsubsec:LoptLx}) luminosities of AT2019wey. These properties, combined with the hard X-ray spectrum reported in Paper I, indicate that AT2019wey is likely a BH system.

Multi-wavelength evolution of \target can be separated into five distinct stages, as illustrated in Figure~\ref{fig:multiwave_lc}. In the dim LHS [i.e., stage (iii)], the UV/optical emission comes from intrinsic thermal emission of an accretion disk with $R_{\rm in}> 100 R_{\rm S}$. In the HIMS [i.e., stage (v)], the UV/optical emission comes from reprocessing of X-rays, and the disk truncation radius has moved inward ($R_{\rm in}\sim 15 R_{\rm S}$). The overall SED evolution fits into the picture of a hot accretion flow consisting of an inner ADAF and a truncated disk. This confirms the widely accepted model for short-period BH LMXBs in the hard state.

The optical light curve of AT2019wey is distinguished by its flatness during stages (iii) and (v). 
This is different from the majority of LMXBs and is similar to what was observed during the 12\,yr outburst of Swift\,J1753.5$-$0127 \citep{Shaw2019, Zhang2019}. 
The X-ray light curve is reminiscent of the `flat top' profile in the 1996 outburst of GRO\,J1655$-$40 \citep{Esin2000}.
As noted before \citep{Esin2000, Shaw2019}, the `standstill' outburst is analogous to the Z Cam class of dwarf novae \citep{Osaki1996}.
In such systems, the mass transfer rate ($\dot M_2$) during quiescence is $\lesssim \dot M_{\rm crit}$. Here $\dot M_{\rm crit}$ is the critical mass-transfer rate, above which the disk remains stable \citep{Dubus1999, Lasota2008}. During the outburst, $\dot M_2$ increased to $\gtrsim \dot M_{\rm crit}$, stabilizing the accretion. In AT2019wey, the second stable period in stage (v) indicates a further increase of $\dot M_2$, probably caused by irradiation on the accretion disk or the companion star.

We note that if AT2019wey continues to remain sufficiently bright in the optical for an extended period of time, the next data release of the \gaia mission may help further constrain the distance. Once the distance is settled, future studies can estimate $\dot M_2$ during the stable stages. Comparison between $\dot M_2$ and $\dot M_{\rm crit}$ can provide a key probe to the evolution of X-ray binaries. 

As discussed in Paper I, \srg is sensitive to the population of BH LMXBs with faint X-ray outbursts. These outbursts are generally associated with lower mass accretion rates and shorter orbital periods \citep{Meyer-Hofmeister2004, Wu2010, Tetarenko2016}. The discovery of AT2019wey showcases the possibility of hunting for similar systems in wide-field optical surveys. This has also been demonstrated in the case of the BH LMXB ASASSN-18ey (MAXI\,J1820$+$070), which was first discovered in the optical \citep{Tucker2018}, and then in the X-ray \citep{Kawamuro2018}. Perhaps the easiest approach to identify similar LMXBs is to study optical light curves of \srg point sources in the Galactic plane.

\begin{acknowledgements}
We thank the anonymous reviewer for providing comments that have significantly improved this manuscript. We thank Mark McKinnon and Amy Mioduszewski for allocating DD time on VLA. We thank Jie Lin and Stephen Smartt for helpful comments.

Y.Y. is supported in part by the Heising-Simons Foundation. M.M.K. acknowledges generous support from the David and Lucille Packard Foundation. E.C.B. is supported in part by the NSF AAG grant 1812779 and grant \#2018-0908 from the Heising-Simons Foundation.

This work is based on observations obtained with the 48-inch Samuel Oschin Telescope and the 60-inch Telescope at the Palomar Observatory as part of the Zwicky Transient Facility project. ZTF is supported by the National Science Foundation under grant No. AST-1440341 and a collaboration including Caltech, IPAC, the Weizmann Institute for Science, the Oskar Klein Center at Stockholm University, the University of Maryland, the University of Washington, Deutsches Elektronen-Synchrotron and Humboldt University, Los Alamos National Laboratories, the TANGO Consortium of Taiwan, the University of Wisconsin at Milwaukee, and Lawrence Berkeley National Laboratories. Operations are conducted by COO, IPAC, and UW. 

SED Machine is based upon work supported by the National Science Foundation under grant No. 1106171. This work was supported by the GROWTH project funded by the National Science Foundation under grant No 1545949. The ZTF forced-photometry service was funded under the Heising-Simons Foundation grant \#12540303 (PI: Graham).

This work has made use of data from the Asteroid Terrestrial-impact Last Alert System (ATLAS) project. The Asteroid Terrestrial-impact Last Alert System (ATLAS) project is primarily funded to search for near-Earth asteroids through NASA grants NN12AR55G, 80NSSC18K0284, and 80NSSC18K1575; byproducts of the NEO search include images and catalogs from the survey area. This work was partially funded by Kepler/K2 grant J1944/80NSSC19K0112 and HST GO-15889, and STFC grants ST/T000198/1 and ST/S006109/1. The ATLAS science products have been made possible through the contributions of the University of Hawaii Institute for Astronomy, the Queen’s University Belfast, the Space Telescope Science Institute, the South African Astronomical Observatory, and The Millennium Institute of Astrophysics (MAS), Chile.

We acknowledge ESA \gaia, DPAC, and the Photometric Science Alerts Team (\url{http://gsaweb.ast.cam.ac.uk/alerts}).

\end{acknowledgements}

\software{
\texttt{astropy} \citep{Astropy2013},
\texttt{CASA} (v5.6.1; \citealt{McMullin2007}), 
\texttt{diskir} \citep{Gierlinski2008, Gierlinski2009},
\texttt{emcee} \citep{Foreman-Mackey2013},
\texttt{FPipe} \citep{Fremling2016},
\texttt{HEASoft} (v6.27; \citealt{Heasarc2014}),
\texttt{LPipe} \citep{Perley2019lpipe}, 
\texttt{makee} (\url{https://sites.astro.caltech.edu/~tb/ipac_staff/tab/makee/}),
\texttt{matplotlib} \citep{Hunter2007},
\texttt{pandas} \citep{McKinney2010},
\texttt{pyraf-dbsp pipeline} \citep{Bellm2016}, 
\texttt{scipy} \citep{Virtanen2020},
\texttt{spextool} \citep{Cushing2004}, 
\texttt{XRB-LrLx\_pub} (\citealt{arash_bahramian_2018_1252036}, \url{https://github.com/bersavosh/XRB-LrLx_pub}),
\texttt{xspec} (v12.11.0; \citealt{Arnaud1996})
\texttt{xtellcor} \citep{Vacca2003}
}

\facilities{PO:1.2m (ZTF, iPTF, POSS), PO:1.5m (SEDM), Gaia, Hale (DBSP, CHIMERA), Keck:I (LRIS), Keck:II (ESI, NIRES), VLA, MAXI, Swift (UVOT, XRT), NICER, Sloan, PS1}

\appendix
\section{Archival Limits}
\subsection{Optical Limits} \label{subsec:optical_limit}
We conducted an archival search of optical photometry at the position of \target. The source was not detected by historical optical surveys, including the Palomar Observatory Sky Survey I (POSS-I, \citealt{Minkowski1963}), the Second Palomar Observatory Sky Survey (POSS-II, \citealt{Reid1991}), SDSS, and the Panoramic Survey Telescope and Rapid Response System DR1 (Pan-STARRS, PS1) \citep{Flewelling2020, Waters2020}, the intermediate Palomar Transient Factory (iPTF; \citealt{Law2009, Rau2009}), and the ZTF.
We list 5$\sigma$ upper limits in Table~\ref{tab:optical_upperlim}. 

\begin{deluxetable}{l|c|c|c|c}[htbp!]
	\tablecaption{Historical upper limits at the position of 
	\target.\label{tab:optical_upperlim}}
	\tablehead{
		\colhead{Survey} &
		\colhead{Time} &
		\colhead{Filter } &
		\colhead{$\lambda_{\rm eff}$ (\AA)}&
		\colhead{Limit}  
	}
	\startdata
	POSS-I 	&1953-10-08 & $r$ & $6500$ & 19.5 \\ 
	POSS-II  &1990-10-26 & $r$ & $6500$ & 20.8 \\ 
	\hline
		\multirow{5}{*}{SDSS}	&\multirow{5}{*}{2004-10-15}& $u$ & 3560 & 22.5 \\
	 	&  & $g$ & 4720 & 23.1 \\
 	    &  & $r$ & 6190 & 22.6 \\
 	    &  & $i$ & 7500 & 22.0 \\
 	    &   & $z$ & 8960 & 20.9 \\
	\hline
	\multirow{5}{*}{PS1} & \multirow{5}{*}{2010-02--2014-12} & $g$ & 4870 & 22.7 \\ 
	&& $r$ & 6210 & 22.3 \\ 
	&& $i$ & 7540 & 22.1 \\ 
	&& $z$ & 8680 & 21.8 \\
	&& $y$ & 9630 & 20.8 \\
	\hline
	iPTF    &2014-01-24 & $R$ & $6420$ & 21.0 \\
	\hline
	\multirow{2}{*}{ZTF} & \multirow{2}{*}{2017-12--2019-11} & $g$ & $4810$ & 21.3 \\
	 &  & $r$ & $6420$ & 21.5 \\
	\enddata
\end{deluxetable}

\subsection{Radio Limit} \label{subsec:radio_limit}
\target was not detected in any archival radio database. The NRAO VLA Sky Survey (NVSS, \citealt{Condon1998}) provides an upper limit of 2\,mJy at 1.4\,GHz 
in 1993--1996. The Karl G. Jansky Very Large Array Sky Survey (VLASS, \citealt{Lacy2020}) provides a 3$\sigma$ upper limit of 0.40\,mJy at 2--4\,GHz in March 2019.

\begin{deluxetable}{ccccc}[htbp!]
	\tablecaption{ZTF Forced Photometry of \target.\label{tab:phot}}
	\tablehead{
		\colhead{MJD} &
		\colhead{$f_\nu$ ($\mu$Jy)} &
		\colhead{$\sigma_{f_\nu}$ ($\mu$Jy)} &
		\colhead{Filter} &
	}
	\startdata
58206.1662  & -12.13  & 7.82 & $g$ \\
58207.1664  & 0.03  & 9.18 & $g$ \\
58210.2064  & 3.97  & 12.89 & $g$ \\
58218.2068  & 3.30  & 7.63 & $r$ \\
58219.1712  & -2.47  & 10.48 & $r$ \\
58231.1454  & -11.46  & 7.32 & $r$ \\
58234.1575  & -7.03  & 9.70 & $g$ \\
58236.1591  & -0.51  & 12.99 & $g$ \\
		\enddata
	\tablecomments{Data up to 2020 November 30 is included. $f_\nu$ is observed flux density (without extinction correction). (This table is available in its entirety in machine-readable form.)}
\end{deluxetable}

\begin{deluxetable}{ccccc}[htbp!]
	\tablecaption{UVOT and SEDM photometry of \target.  \label{tab:uvot}}
	\tablehead{
		\colhead{Date}   
		& \colhead{Instrument}  
		& \colhead{Filter}  
		& \colhead{$m$} 
	}
	\startdata
	2020-04 Coadd & \swift/UVOT &  $ B $ & $18.93 \pm 0.17$ \\
    2020-04 Coadd & \swift/UVOT &  $ U $ & $20.16 \pm 0.24$ \\
    2020-04 Coadd & \swift/UVOT &  $ uvm2 $ & $22.55 \pm 0.42$ \\
    2020-04 Coadd & \swift/UVOT &  $ uvw1 $ & $21.17 \pm 0.27$ \\
    2020-04 Coadd & \swift/UVOT &  $ uvw2 $ & $22.86 \pm 0.50$ \\
    2020-04 Coadd & \swift/UVOT &  $ V $ & $18.00 \pm 0.15$ \\
    2020-08-05 & \swift/UVOT &  $ uvm2 $ & $> 21.16$ \\
    2020-08-09 & \swift/UVOT &  $ uvm2 $ & $22.16 \pm 0.33$ \\
    2020-08-12 & \swift/UVOT &  $ uvw2 $ & $21.83 \pm 0.21$ \\
    2020-08-19 & \swift/UVOT &  $ U $ & $19.35 \pm 0.06$ \\
    2020-08-26 & \swift/UVOT &  $ uvw1 $ & $20.78 \pm 0.13$ \\
    2020-09-02 & \swift/UVOT &  $ uvm2 $ & $22.12 \pm 0.43$ \\
    2020-09-09 & \swift/UVOT &  $ uvw2 $ & $22.00 \pm 0.24$ \\
    2020-09-16 & \swift/UVOT &  $ U $ & $19.26 \pm 0.07$ \\
    2020-09-23 & \swift/UVOT &  $ uvw1 $ & $20.72 \pm 0.13$ \\
	\hline
	2020-10-21 & P60/SEDM & $U$ & $19.11\pm 0.09$ \\
	2020-10-25 & P60/SEDM & $U$ & $19.21\pm 0.31$ \\
	\enddata
	\tablecomments{$m$ is observed magnitude (without extinction correction).}
\end{deluxetable}

\section{Instrumental/Observational Information} \label{sec:obsinfo}

We provide ZTF photometry in Table~\ref{tab:phot}. UVOT and SEDM photometry is provided in Table~\ref{tab:uvot}.

We obtained optical spectroscopic follow-up observations of \target using the Low Resolution ($R\approx 1000$) Imaging Spectrograph (LRIS; \citealt{Oke1995}) on the Keck-I telescope, the Double Spectrograph (DBSP; $R\approx 1200$; \citealt{Oke1982}) on the 200-inch Hale telescope, and the medium-resolution ($R\approx13000$) Echellette Spectrograph and Imager (ESI; \citealt{Sheinis2002}) on the Keck-II telescope. We obtained NIR spectroscopy using the Near infrared emission spectroscopy (NIRES; $R=2700$) on the Keck-II telescope. Spectroscopic observations were coordinated with the GROWTH Marshal \citep{Kasliwal2019}.

The DBSP spectra were reduced using the \texttt{pyraf-dbsp} pipeline \citep{Bellm2016}. The LRIS spectra were reduced and extracted using \texttt{Lpipe} \citep{Perley2019lpipe}. The flat-fielding, wavelength solution (using sky lines) and extraction for the NIRES spectrum was carried out using the \texttt{spextool} code \citep{Cushing2004}. The extracted spectrum was flux calibrated using the telluric A0V standard star HIP\,16652 with the \texttt{xtellcor} code \citep{Vacca2003}. The ESI spectrum was reduced using the \texttt{MAKEE}\footnote{\url{http://www.astro.caltech.edu/~tb/ipac_staff/tab/makee/}} pipeline following the standard procedure. Flux calibration was not performed on the ESI spectrum.

\section{Details of Analysis}
\begin{figure}[htbp!]
	\centering
	\includegraphics[width=\columnwidth]{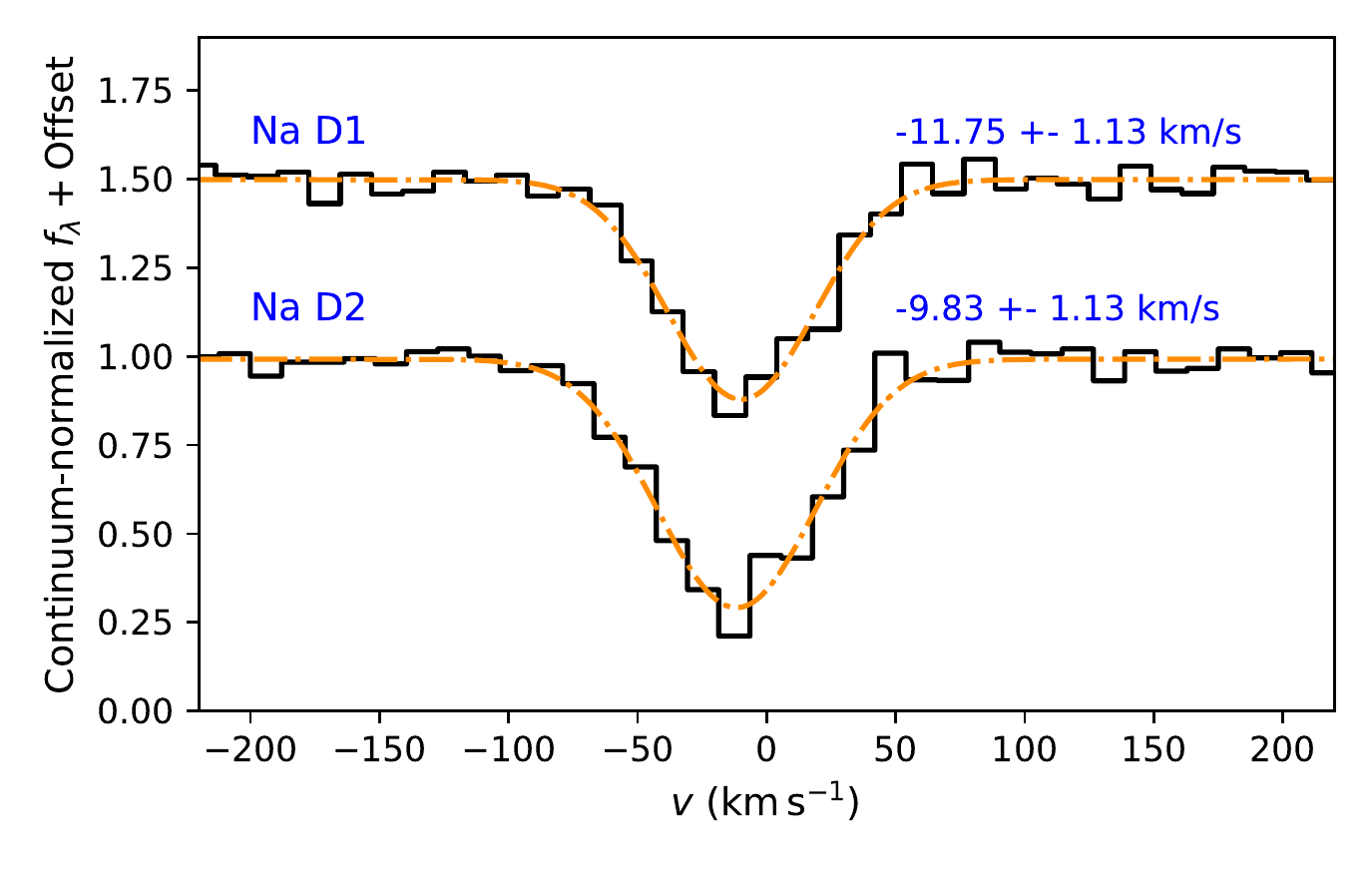}
	\caption{\ion{Na}{I} D lines in velocity space fitted with a Gaussian (dash-dotted lines). The spectrum is heliocentric velocity corrected. \label{fig:esi}}
\end{figure}
\subsection{Extinction Estimation}\label{subsec:extinction}

The $EW$ of interstellar absorption lines has been observed to be correlated with the amount of reddening. To estimate the extinction of \target, we produced a summed spectrum from the LRIS and ESI spectra. We did not include DBSP spectra in this analysis since the CCD malfunction resulted in non-astrophysical structures between 5750\,\AA\ and 6200\,\AA\ in the continuum. This problem prevents $EW$ of spectral lines from being accurately determined from DBSP spectra. The $EW$ of DIB $\lambda \lambda5780$, $\lambda6283$ and \ion{Na}{I} D lines were measured from the summed spectrum. 
As a result, we got $EW(\lambda 5780) =0.56\pm 0.02\,{\rm \AA}$, and $EW(\lambda 6283) =1.55 \pm 0.02\,{\rm \AA}$. These can be converted to $E(B-V) = 0.92 \pm 0.02$ and $1.23\pm 0.02$ using relations presented by \citet{Yuan2012}. We got $EW($\ion{Na}{I} D$)= 1.84\pm0.02\,{\rm \AA}$, which can be converted to $E(B-V) = 2.01 \pm 0.38$ using the relation in \citet{Poznanski2012}.

The inferred $E(B-V)$ values are greater than the total Galactic extinction of $E(B-V)=0.88$ \citep{Schlafly2011}. However, we note that at the measured $EW$, the calibration uncertainty is large. From \citet[upper panels of Fig.~4]{Yuan2012} and \citet[bottom panel of Fig.~9]{Poznanski2012}, we infer that $E(B-V)$ towards \target should be $\gtrsim 0.8$. 

We also attempt to infer the extinction by assuming that the 6000--10000\,\AA\ March 23 LRIS spectrum is in the Rayleigh--Jeans (R--J) tail of a blackbody ($f_\lambda \propto \lambda^{-4}$ when $h\nu \ll kT$), which yields $E(B-V)=1.29$ and a blackbody radius ($R_{\rm bb}$) of
\begin{align}
    R_{\rm bb} = (4.5\times 10^{10}\,{\rm cm}) \left(\frac{D}{5\,{\rm kpc}}\right)  \left( \frac{T_{\rm bb}}{5.0\times 10^{4}\,{\rm K}}\right)^{-1/2}
\end{align}
Note that this is likely an overestimate of the true extinction (and a lower limit of the outer disk radius), since the optical is only in the R--J limit when $kT \gg 2$\,eV ($T \gg 2\times 10^4$\,K). For instance, for an extinction of $E(B-V)\sim 0.9$, we have 
\begin{align}
    R_{\rm bb} = (1.0\times 10^{11}\,{\rm cm}) \left(\frac{D}{5\,{\rm kpc}}\right)  \left( \frac{T_{\rm bb}}{1.1\times 10^{4}\,{\rm K}}\right)^{-1/2}
\end{align}

\subsection{Lower Limit of Distance}\label{subsec:distance}

In Appendix~\ref{subsec:extinction}, we find that \target should have an extinction of $0.8 \lesssim E(B-V) \lesssim 1.2$. If this is from diffuse interstellar absorption, the distance of \target should be greater than 1\,kpc using the map of STructuring by Inversion the Local Interstellar Medium (Stilism\footnote{\url{https://stilism.obspm.fr/}}; \citealt{Capitanio2017}). 

We are able to put a lower limit to the distance using the velocity of the \ion{Na}{I} D doublets in the ESI spectrum, given that the lines arise from interstellar absorption by a dust cloud along the line-of-sight to \target. The velocities of D1 and D2 lines were measured to be $-11.75\pm1.13\,{\rm km\,s^{-1}}$ and $-9.83\pm1.13\,{\rm km\,s^{-1}}$, respectively (see Figure~\ref{fig:esi}). Assuming that the velocity of the dust cloud follows Galactic rotation, we have
\begin{align}
V_{\rm obs, r} = A d {\rm sin}(2l) \label{eq:oortA}
\end{align}
where $A=15.3\pm 0.4\,{\rm km\, s^{-1}\, kpc^{-1}}$ is the Oort $A$ constant \citep{Bovy2017}, $l= 151.2^{\circ}$ is the Galactic longitude of \target, and $d$ is distance to the dust cloud. Therefore, Eq.~(\ref{eq:oortA}) gives $d=0.83$\,kpc.

\bibliography{at2019wey}{}
\bibliographystyle{aasjournal}

\end{document}